\documentclass[11pt]{article}

\usepackage{graphicx, amsfonts, amsbsy, amssymb, amsthm, amsmath}
\usepackage{bbm}
\usepackage{tikz}

\usepackage{authblk}

\renewcommand{\vec}[1]{{\mathbf{#1}}}

\renewcommand{\Re}{\mathop{\mathfrak{Re}}}
\renewcommand{\Im}{\mathop{\mathfrak{Im}}}

\newcommand{\rmd}{{\mathrm d}}
\newcommand{\rme}{{\mathrm e}}
\newcommand{\rmi}{{\mathrm i}}

\newcommand{\Ord}{{\mathrm O}}

\DeclareSymbolFont{lettersA}{U}{pxmia}{m}{it}

\DeclareMathSymbol{\alphaup}{\mathord}{lettersA}{"0B}
\DeclareMathSymbol{\betaup}{\mathord}{lettersA}{"0C}
\DeclareMathSymbol{\gammaup}{\mathord}{lettersA}{"0D}
\DeclareMathSymbol{\deltaup}{\mathord}{lettersA}{"0E}
\DeclareMathSymbol{\epsilonup}{\mathord}{lettersA}{"22}
\DeclareMathSymbol{\zetaup}{\mathord}{lettersA}{"10}
\DeclareMathSymbol{\etaup}{\mathord}{lettersA}{"11}
\DeclareMathSymbol{\thetaup}{\mathord}{lettersA}{"12}
\DeclareMathSymbol{\iotaup}{\mathord}{lettersA}{"13}
\DeclareMathSymbol{\kappaup}{\mathord}{lettersA}{"14}
\DeclareMathSymbol{\lambdaup}{\mathord}{lettersA}{"15}
\DeclareMathSymbol{\muup}{\mathord}{lettersA}{"16}
\DeclareMathSymbol{\nuup}{\mathord}{lettersA}{"17}
\DeclareMathSymbol{\xiup}{\mathord}{lettersA}{"18}
\DeclareMathSymbol{\piup}{\mathord}{lettersA}{"19}
\DeclareMathSymbol{\rhoup}{\mathord}{lettersA}{"1A}
\DeclareMathSymbol{\sigmaup}{\mathord}{lettersA}{"1B}
\DeclareMathSymbol{\tauup}{\mathord}{lettersA}{"1C}
\DeclareMathSymbol{\upsilonup}{\mathord}{lettersA}{"1D}
\DeclareMathSymbol{\phiup}{\mathord}{lettersA}{"1E}
\DeclareMathSymbol{\chiup}{\mathord}{lettersA}{"1F}
\DeclareMathSymbol{\psiup}{\mathord}{lettersA}{"20}
\DeclareMathSymbol{\omegaup}{\mathord}{lettersA}{"21}

\newcommand{\vecalpha}{{\pmb{\alphaup}}}
\newcommand{\vecbeta}{{\pmb{\betaup}}}

\renewcommand{\Psi}{\varPsi}
\renewcommand{\Lambda}{\varLambda}
\renewcommand{\Sigma}{\varSigma}
\renewcommand{\Gamma}{\varGamma}
\renewcommand{\Theta}{\varTheta}
\renewcommand{\Xi}{\varXi}
\renewcommand{\Pi}{\varPi}
\renewcommand{\Upsilon}{\varUpsilon}
\renewcommand{\Phi}{\varPhi}
\renewcommand{\Omega}{\varOmega}

\newcommand{\R}{{\mathbb R}}

\newcommand{\Q}{{\mathbb Q}}

\newcommand{\T}{{\mathbb T}}
\newcommand{\C}{{\mathbb C}}

\newcommand{\cosec}{\mathop{\mathrm{cosec}}\nolimits}

\newcommand{\coloneq}{\mathbin{\hbox{\raise0.08ex\hbox{\rm :}}\!\!=}}
\newcommand{\eqcolon}{\mathbin{=\!\!\hbox{\raise0.08ex\hbox{\rm :}}}}

\renewcommand{\leq}{\leqslant}
\renewcommand{\geq}{\geqslant}
\renewcommand{\epsilon}{\varepsilon} 

\newcommand{\dimostrazione}{\noindent{\sl Proof.}\phantom{X}}
\newcommand{\dimostrazionea}[1]{\noindent{\sl Proof of #1.}\phantom{X}}
\newcommand{\finire}{\hspace*{\fill}~$\Box$}

\newcommand \printdate[3]{%
    \def \@suffix##1{%
        \def \@n{##1}%
        \ifnum \@n = 1 st\else%
        \ifnum \@n = 2 nd\else%
        \ifnum \@n = 3 rd\else%
        \ifnum \@n = 21 st\else%
        \ifnum \@n = 22 nd\else%
        \ifnum \@n = 23 rd\else%
        \ifnum \@n = 31 st\else%
        th\fi \fi \fi \fi \fi \fi \fi%
    }%
    \relax%
    \number #1\raise0.7ex\hbox{\footnotesize \@suffix{#1}}\kern0.25em%
    \ifcase #2\or%
        January\or February\or March\or%
        April\or May\or June\or%
        July\or August\or September\or%
        October\or November\or December%
    \fi\ %
    \number #3%
}

\newtheorem{theorem}{Theorem}[section]
\newtheorem{proposition}[theorem]{Proposition}

\newtheorem{lemma}[theorem]{Lemma}

\theoremstyle{definition}
\newtheorem{remark}[theorem]{Remark}

\DeclareMathOperator{\diag}{diag}
\DeclareMathOperator{\logarithm}{log}
\DeclareMathOperator{\Eig}{Eig}

\renewcommand{\log}{\logarithm}  %
\renewcommand{\ln}{\logarithm}   
\newcommand{\Ent}{{\mathsf{S}}}
\newcommand{\ent}{{\mathsf{s}}}
\newcommand{\REnt}{{\mathsf{R}}}
\newcommand{\rent}{{\mathsf{r}}}
\newcommand{\ain}{a^{\mathrm{in}}}
\newcommand{\aout}{a^{\mathrm{out}}}

\renewcommand{\vec}[1]{\boldsymbol{#1}}  
\newcommand{\curlyF}{{\mathcal{F}}}
\newcommand{\wplus}{\vec{w}_+}
\newcommand{\wminus}{\vec{w}_-}
\newcommand{\wplusi}{\vec{w}_{+\rmi}}
\newcommand{\wminusi}{\vec{w}_{-\rmi}}
\newcommand{\aplus}{\alpha_+}
\newcommand{\aminus}{\alpha_-}
\newcommand{\aplusi}{\alpha_{+\rmi}}
\newcommand{\aminusi}{\alpha_{-\rmi}}
\numberwithin{equation}{section}

\usepackage[text={15cm,23cm}]{geometry} 

\usepackage[colorlinks=true,%
            citecolor=blue,%
            urlcolor=darkgray]{hyperref}  

\begin{document}
\title{Maximal scarring for eigenfunctions of quantum graphs}
\author{G.~Berkolaiko}
\affil{Department of Mathematics, Texas A\&M University, College Station, 
TX77843-3368, U.S.A.}
\author{B.~Winn}
\affil{Department of Mathematical Sciences, 
Loughborough University, Loughborough,
LE11 3TU, U.K.}
\date{\printdate{25}{6}{2018}}
\maketitle
\begin{abstract}
  We prove the existence of scarred eigenstates for star graphs with
  scattering matrices at the central vertex which are either a Fourier
  transform matrix, or a matrix that prohibits back-scattering. We
  prove the existence of scars that are half-delocalised on a single
  bond.  Moreover we show that the scarred states we construct are
  maximal in the sense that it is impossible to have quantum
  eigenfunctions with a significantly lower entropy than our examples.

  These scarred eigenstates are on graphs that exhibit generic
  spectral statistics of random matrix type in the large graph limit,
  and, in contrast to other constructions, correspond to
  non-degenerate eigenvalues; they exist for almost all choices of
  lengths.
\end{abstract}

\section{Introduction}
\label{sec:intro}
The possibility of existence of scarred quantum eigenstates has been a
mystery of quantum mechanics since the intriguing suggestion of Heller
\cite{hel:bse} that for a quantum Hamiltonian corresponding to a
chaotic classical limit, subsequences of eigenfunctions can
converge in the high energy limit to a measure supported on one-or-more
short unstable periodic orbits.  This is the ``strong'' notion of
scarring, as compared to the phenomenon of weak scarring, in which
states are averaged over energy windows that are semiclassically increasing 
in size and which is now well-understood 
\cite{bog:swf,aga:scf,ber:qso,kap:siq,kea:oba,sch:soq}.

Mathematically rigorous constructions of quantum scars have been found
in some model systems: for quantum cat maps \cite{fau:sef}, and for
some families of quantum graphs \cite{cdv:scm,ber:nqe}, the topic to
which the present article is devoted. (See also
\cite{kel:soi,ana:eos,has:ebt} for results pertaining to
eigenfunctions which localise around other invariant structures.)  In
the opposite direction, in certain situations it has been rigorously
proved that scars cannot exist.  These are models for which the
quantum unique ergodicity property holds \cite{lin:ima,bro:gea}.

In one sense, the question of ``perfect'' scarring as strictly
described above has also been settled for a broad class of systems.
On compact Riemann surfaces whose geodesic flow is chaotic, 
the entropy of quantum limits of eigenfunctions has a strictly
positive lower bound \cite{ana:eat,ana:hdo,riv:eos}. A limiting measure 
supported on a finite union of periodic orbits would have entropy
zero, and is thus ruled-out.  However, the possibility remains
open of a quantum limit that has a  positive \emph{proportion} of its mass
supported on a periodic orbit, with the remaining mass, say equi-distributed.
For a surface of constant negative curvature the results of \cite{ana:hdo}
put a lower bound of $\frac12$ on the entropy of quantum limits,
meaning that at most half the mass can be carried by periodic orbits.
The scars constructed on cat maps in \cite{fau:sef} are also 
half-localised on periodic orbits and half equi-distributed, which
was proved to be the maximal amount of delocalisation possible in 
that context in \cite{fau:otm}.

The study of statistical properties of quantum eigenfunctions and
eigenvalues, particularly when the underlying classical system
is chaotic, is part of the field of quantum chaos.  The study of
Schr\"odinger operators on one-dimensional networks is a prominent
part of this field, going by the name of `quantum graphs'
\cite{kot:qco,kot:pot,ber:tps,kot:qga,bar:otl,ber:sga,ber:tlo,bol:tsc,
ber:fffa,ber:fffl,gnu:sco,gnu:qeo,cdv:scm,bra:qef,kam:eof} (the
list of references is highly-incomplete but gives a flavour of the
subject).  In this article
we report on constructions of scarred eigenfunctions of quantum 
graphs, that in some cases exhibit maximal delocalisation.
We defer a full introduction to the quantum graphs to section
\ref{sec:two}, mentioning here only the necessary facts to introduce our
results.

A graph is a network of vertices and bonds on which waves propagate
and are scattered in accordance with scattering matrices attached to
each vertex.  An eigenfunction is a standing wave, which may be
completely described by $2B$-dimensional vector $\vec{a}$ giving the
complex amplitudes of the wave on each \emph{directed} bond.  We will
be interested in subsequences of eigenvectors converging to a limit
that shows \emph{localisation}---significant enhanced amplitude on
certain subsets of bonds.

Colin de Verdi\`ere \cite{cdv:scm} has studied scarred eigenvectors of
quantum graphs with Kirchhoff scattering matrices at all vertices, and
proved that for irrational and independent bond lengths there exist
subsequences of eigenfunctions which fully localise onto simple paths
that are either: closed paths, or connect two distinct vertices of
degree one (the former was initially observed in \cite{sch:soq}, the
latter may be seen as a significant generalisation of an earlier
result \cite{ber:nqe} by the authors).  In \cite{cdv:scm} a
convergence result is proved that will be a key tool in our analyses
(see theorem \ref{thm:cdv} below).

Our main results concern star graphs, but we will
focus on alternative choices of boundary conditions.  Star graphs are
graphs with a single central vertex surrounded by $B$ outlying
vertices (see figure \ref{fig:star_graph} below).  It has been known
for some time \cite{ber:iws,ber:nqe} that there exist subsequences of
eigenfunctions that localise on a pair of bonds for quantum star
graphs with irrational bond lengths and Kirchhoff boundary conditions
at the central vertex.  We study certain non-Kirchhoff boundary
conditions that have attracted interest in the
subject\footnote{Principally because, in constrast to Kirchhoff star
  graphs, they exhibit spectral statistics agreeing with Random Matrix
  Theory; see the discussion below.} and prove that there exist
subsequences of eigenfunctions that half-localise on a single bond.

By `half-localise' we mean that there exist limits of eigenfunctions
which are an equal superposition of a state with all mass 
equally-distributed to all bonds, and a state that is fully
concentrated on a single bond.  These eigenfunctions are
maximally scarred, as measured by their entropy, a fact which we show
in section \ref{sec:five}.  

The boundary conditions that we consider for the star graphs are
those given by the Fourier transform matrix, and equi-transmitting
matrices \cite{har:qgw}.  Both kinds of matrices are unitary; 
Fourier transform matrices have all components with equal amplitude,
and equi-transmitting matrices have all off-diagonal components having
equal amplitudes and diagonal entries zero.

The $6\times 6$ Fourier transform matrix and an example of a $6\times6$
equi-transmitting matrix are
\begin{equation}
  \label{eq:89}
  {\mathcal{F}}_6 = \frac1{\sqrt{6}}\left( \begin{array}{cccccc}
                  1 & 1 & 1 & 1 & 1 & 1 \\
                  1 & \omega & \omega^2 & \omega^3 & \omega^4 & \omega^5 \\
                  1 & \omega^2 & \omega^4 & 1 & \omega^2 & \omega^4 \\
                  1 & \omega^3 & 1 & \omega^3 & 1 & \omega^3 \\
                  1 & \omega^4 & \omega^2 & 1 & \omega^4 & \omega^2 \\
                  1 & \omega^5 & \omega^4 & \omega^3 & \omega^2 & 1 
                           \end{array}\right),\quad
  E_6 = \frac{1}{\sqrt{5}} \left( \begin{array}{crrrrr}
                                    0 & 1 & 1 & 1 & 1 & 1 \\
                                    1 & 0 & 1 & -1 & -1 & 1 \\
                                    1 & 1 & 0 & 1 & -1 & -1 \\
                                    1 & -1 & 1 & 0 & 1 & -1 \\
                                    1 & -1 & -1 & 1 & 0 & 1 \\
                                    1 & 1 & -1 & -1 & 1 & 0 
                                  \end{array} \right),
\end{equation}
with $\omega=\rme^{\rmi\pi/3}$.  Fourier transform matrices exist in any
size, but it is not known whether or not $n\times n$ equi-transmitting 
matrices exist for every $n>3$ (it is known that no $3\times3$ 
equi-transmitting
matrix exists).  For example the authors are unaware of the existence
of a $7\times 7$ equi-transmitting matrix.

Our motivation for considering star graphs quantised with these
scattering matrices comes from a consideration of the statistics
of eigenvalues.  For a quantum system arising as the quantisation
of a classically chaotic system, the Random Matrix conjecture
\cite{boh:coc,cas:otc} states that generically the eigenvalues
do not occur independently, but rather are correlated at scales of
the size of the mean spacing in exactly
the same way as the limit in large matrix size of eigenvalues
of ensembles of random matrices.  The appropriate ensemble is
governed by broad properties of the system such as whether it
exhibits time-reversal invariance.

In earlier cases where eigenfunctions of quantum systems are
rigorously know to scar \cite{fau:sef,ber:nqe}, the corresponding
spectral correlations are different from those predicted by Random
Matrix Theory (see \cite{kea:tcm} and \cite{ber:tps,ber:sga}).  This
does not invalidate the Random Matrix conjecture; it simply makes
those systems non-generic.  Also, the construction in \cite{fau:sef}
makes essential use of the fact that the corresponding eigenspaces 
are highly degenerate, so the fact that our eigenstates are simple
is notable.

Star graphs with Fourier transform scattering matrices and
equi-transmitting scattering matrices are expected to give spectral
statistics consistent with the predictions of Random Matrix Theory
\cite{gnu:qga} as the size of the graph tends to infinity.  Our
results, together with those of \cite{cdv:scm} establish the existence
of scarred eigenfunctions in systems where (in the limit of large
graph size) the spectral statistics seem to be generic for chaotic
systems.  

Our results presented in sections \ref{sec:five}--\ref{sec:equi-transmit}
involve analyses of rank-one perturbations of certain complex Hadamard
matrices and complex conference matrices, which may be of independent
interest.

\section{Quantum graphs}
\label{sec:two}

We introduce, in a brief way, the main definitions and ideas of
quantum graphs. For details not given here we refer to \cite{ber:itq},
the review articles \cite{kuc:qgI,gnu:qga} and 
the seminal papers \cite{kot:qco,kot:pot} on quantum chaos on graphs.

\begin{figure}
  \centering
  \begin{tikzpicture}
\begin{scope}[very thick, 
              every node/.style={fill, circle, inner sep=3pt, outer sep=0pt}]
    \node (v) at (0,0) {};
    \draw (v) -- (200:2.5) node[label={200:$1$}] {};
    \draw (v) -- (250:2) node[label={200:$B$}] {};
    \draw (v) -- (130:1.4) node[label={130:$2$}] {};
    \draw (v) -- (70:2.1) node[label={70:$3$}] {};
    \draw (v) -- (0:1.9) node[label={0:$4$}] {};
  \end{scope}
  \node[rotate=90] at (300:1.1) {$\ddots$};
  \end{tikzpicture}
  \caption{A star graph with $B$ bonds}
  \label{fig:star_graph}
\end{figure}
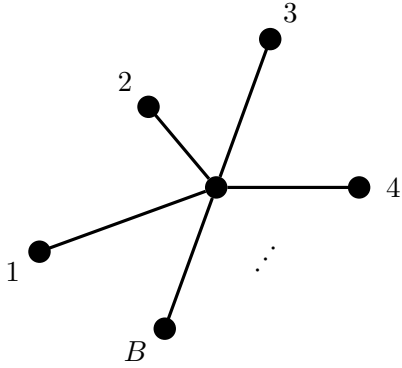

A graph $\Gamma=(V,E)$ consists of a set of vertices $V$ and bonds (or
edges) $E$ with bonds connecting pairs of vertices.  
Our results are for star graphs.  These are
graphs with $B$ bonds each connecting outlying vertices to a single
central vertex (see figure \ref{fig:star_graph}).

We make a graph $\Gamma$ into a metric graph by associating a positive
length $L_b$ to each bond $b\in E$. It is convenient to record the
bond lengths in a vector $L=(L_b)_{b\in E}\in\R_{>}^B$, where $B$ is
the number of bonds.  In many applications, and indeed sometimes
below, the bond lengths $L$ will be chosen to be irrational and
linearly independent over $\Q$.

A metric graph is made into a \emph{quantum graph} by one of two
procedures.  The first such procedure specifies a differential
operator---usually of Schr\"odinger type---on the edges of the
graph and matching conditions at the vertices.  The second procedure,
more popular in physical applications \cite{cha:pqt,sch:ssf,tan:usm},
is to consider free wave propagation on the edges of the graph and
scattering at the vertices.  The two procedures are not
entirely equivalent;  an interested reader can find a comparison in
\cite{ber:tco}. We also note here that the second (``scattering''
or ``wave-dynamics'') procedure is equivalent to the spectral problem
for a first-order differential operator on a directified graph
\cite{car:iep}.

We will not describe the wave-dynamics model of quantum graphs in
detail.  The wave-function on a bond $b = (u,v)$ (here $u$ and $v$
are some vertices) is taken to be a superposition of two plain waves
with momentum $k$, one propagating from $u$ to $v$ with the amplitude
$a_{u,v}$ (as measured at $v$) and the other propagating from $v$ to
$u$ with the amplitude $a_{v,u}$ (as measured at $u$),
\begin{equation}
  \label{eq:108}
  \psi_{u,v}(x) = a_{u,v} \rme^{-\rmi k (L_{u,v}-x)} 
  + a_{v,u}\rme^{-\rmi k x}.
\end{equation}
It is convenient to index the amplitudes on the entire graph using
$2B$ \emph{directed} bonds, denoted by $[u,v]$.  A \emph{reversal} operation
is defined naturally on directed bonds by $\overline{[u,v]} =
[v,u]$.

Upon arrival at the vertex $v$, the wave is scattered into several
waves coming out of $v$ and towards vertices $u'$ adjacent to $v$.
Some part of the wave may be reflected back to $u$; this is termed
\emph{back-scattering}.  Conversely, a wave coming out of the vertex $v$
along the bond $[v,u]$ is the superposition of scattered waves which
arrived at the vertex $v$ along bonds $[u',v]$.  This process is
described by a $d\times d$ unitary \emph{vertex scattering} matrix
$\sigma^{v}$ \textit{via} the mapping
\newcommand\mapsfrom{\mathrel{\reflectbox{\ensuremath{\mapsto}}}}
\begin{equation}
  \label{eq:scattering_described}
  a_{v,u} \mapsfrom \rme^{\rmi k L_{v,u}} \sum_{u' \sim v} \sigma^{v}(u,u') 
  \,  a_{u',v},
\end{equation}
where $d$ is the degree of the vertex $v$.  Here the exponential factor
is the phase acquired by the wave while traversing the bond $[v,u]$ of
length $L_{v,u}$.  A value $k$ is an eigenvalue if there is a standing
wave: the mapping \eqref{eq:scattering_described} becomes an equality
for each $[v,u]$.

An example of a scattering matrix for a vertex of
degree $d$, is the (Neumann-)Kirchhoff scattering matrix
\begin{equation}
  \label{eq:neumann_scatter}
  \sigma_{\mathrm N}(u,u') \coloneq \frac2d-\delta_{uu'}.
\end{equation}
This particular scattering matrix may also be obtained through the
first approach for the negative Laplacian with Kirchhoff (or
``natural'') vertex conditions.  An advantage of the wave-dynamics
approach is that one can specify scattering matrices that possess
mathematically appealing properties, even if they do not arise out of
boundary conditions for a self-adjoint operator, as was done in
\cite{sch:ssf, tan:usm, har:qgw} and as we will do below.

To record the scattering matrices for all vertices of a graph, it is
typical to define a $2B\times 2B$ unitary \emph{bond scattering}
matrix $S$ with entries $S_{[v',u'],[u,v]}$ indexed by directed bonds,
\begin{equation}
  \label{eq:bigS_def}
  S_{[v',u'],[u,v]} \coloneq \delta_{vv'}\, \sigma^v(u',u).
\end{equation}
The only non-zero entries of the matrix $S$ are the ones where $v'=v$
and therefore the directed bond $[v',u']$ follows $[u,v]$ according to
the connectivity of the graph.

The matrix $S$ is useful for more than just bookkeeping; it is the key 
to a powerful method of studying spectral properties of the graph
which we now describe.  Define the $2B\times2B$ diagonal matrix $D
= D(k) = D(k,L)$ by
\begin{equation}
  \label{eq:105}
   D_{[u',v'],[u,v]} \coloneq \delta_{uu'}\delta_{vv'} \rme^{\rmi k L_{v,u}}.
\end{equation}
The condition for a standing wave with momentum $k$ (cf.\
\eqref{eq:scattering_described}) takes the form
\begin{equation}
  \label{eq:104}
  \det(I-D(k)S) = 0.
\end{equation}

Assuming that $k=k_n$ is a value such that the determinant in
\eqref{eq:104} vanishes, there is at least one non-zero vector 
$\vec{a}=(a_b)\in\C^{2B}$ such that
\begin{equation}
  \label{eq:106}
  (I-D(k_n)S)\vec{a} = 0,
\end{equation}
and the vector $\vec{a}$ is exactly the vector of amplitudes for the
wave-function \eqref{eq:108}.  We shall refer to vectors $\vec{a}$
satisfying \eqref{eq:106} as quantum graph eigenvectors.  It is common
to call the product $U=D(k,L)S$ the \emph{quantum evolution operator}
for the graph, and $S$ the \emph{bond-scattering matrix} (or
\emph{$S$-matrix}).

Note that writing down matrices $S$ in practice requires an ordering
of the directed bonds.  For star graphs we will always first list the
bonds going to the central vertex (numbered by $j$, $j=1,\ldots,B$)
and then list the bonds going in the opposite direction (numbered
$j+B$, $j=1,\ldots,B$).  Thus, the bond $j+B$ is the reversal of the
bond $j$ and vice versa.  From now on we will use this numbering in
\textit{lieu} of the directed bond notation $[u,v]$. 

In order to prove existence of subsequences of eigenvectors converging
to scarred states, we use the following result of Colin de Verdi\`ere:
\begin{theorem}
  \label{thm:cdv}
Consider a quantum graph with bond scattering matrix $S$.
If $\vec{s}\in\C^{2B}$, $\|\vec{s}\|=1$ is an eigenvector of $D(k_0,L_0)S$ 
for some $k_0$ and some set of bond lengths $L_0$, with a simple 
eigenvalue, then for any choice of incommensurate bond 
lengths $L$, there is a subsequence $(k_{n_j})
\subseteq(k_n)$ such that there exist normalised quantum graph eigenvectors 
$\vec{a}_{n_j}$ with eigenvalue $k_{n_j}$ that satisfy
\begin{equation}
  \label{eq:1}
  \vec{a}_{n_j} \to \vec{s}\qquad\text{as $j\to\infty$.}
\end{equation}
\end{theorem}
\dimostrazione  This is basically part 2 of theorem 2.1 of \cite{cdv:scm},
re-written in the language of scattering matrices.  Note that Kirchhoff 
boundary conditions are assumed in the statement of theorem 2.1 of 
\cite{cdv:scm}, but part 3 thereof is the only place where that assumption 
is necessary. 
Strictly speaking, the quoted result requires $\vec{s}$ to be an
eigenvector of the quantum graph, i.e.\ an eigenvector of $D(k_0,L_0)S$
with eigenvalue $1$, but it is easy to see that if there exists a 
choice of bond lengths $L_0$ for which $\vec{s}$ has any other
eigenvalue, by adjusting all lengths by the same amount we find 
an $L_0'$ for which the eigenvalue is indeed $1$.\finire

The main difficulty in applying theorem \ref{thm:cdv} will be to prove
that the eigenvalue of $D(k_0,L_0)S$ is simple.  In the cases that we
consider we will do this by perturbing the bond lengths in an
appropriate way.  

For the sake of concreteness, we impose Kirchhoff conditions at the
external vertices (leaves) of the star graphs, although other conditions
could be specified there without substantially changing the results.
At the central vertex, we choose scattering matrices that are either
a Fourier transform matrix, or an equi-transmitting matrix.

To measure the degree to which limits of eigenvectors are scarred we will
adopt the entropy measure as proposed on quantum graphs in \cite{kam:eof}.

Define the Shannon entropy $\Ent: \C^{2B}\to[0,\log 2B]$ by
\begin{equation}
\label{eq:Entropy}
  \Ent(\vec{a}) \coloneq -\sum_{b=1}^{2B} \frac{|a_b|^2}{\|\vec{a}\|^2} \ln
\left( \frac{|a_b|^2}{\|\vec{a}\|^2} \right),
\end{equation}
and the R\'enyi entropy $\REnt_\rho: \C^{2B}\to[0,\log 2B]$ by
\begin{equation}
  \label{eq:135}
  \REnt_\rho(\vec{a}) \coloneq -\frac1\rho\ln\Bigg( \sum_{b=1}^{2B}
  \left( \frac{|a_b|^2}{\|\vec{a}\|^2} \right)^{1+\rho}\Bigg),
\end{equation}
for a parameter $\rho>-1$, $\rho\neq0$.  We also define
\begin{equation}
  \label{eq:81}
  \REnt_\infty(\vec{a}) \coloneq -\log \left( \max_{1\leq b\leq 2B}\left\{ 
\frac{|a_b|^2}{\|\vec{a}\|^2} \right\} \right),
\end{equation}
which agrees with taking the limit $\rho\to\infty$ in
\eqref{eq:135}. The R\'enyi entropy generalises the Shannon entropy in
the sense that for $\vec{a}\in\C^{2B}$,
\begin{equation}
  \label{eq:136}
  \lim_{\rho\to0} \REnt_\rho(\vec{a}) = \Ent(\vec{a}).
\end{equation}

For a completely delocalised vector $\vec{a}=\frac1{\sqrt{2B}}(1,\ldots,1)$
we have $\Ent(\vec{a})=\REnt_\rho(\vec{a})=\log 2B$, whereas for an 
eigenvector localised on one single bond, $\Ent(\vec{a})=\REnt_\rho(\vec{a})
=\log2$, so, roughly-speaking, the
smaller the entropy, the greater the measure of localisation, or strength
of the scarring.  On the other hand, the entropic uncertainty principle
\cite{maa:geu} places a theoretical lower-bound on the entropy of 
eigenstates. In section \ref{sec:three_point_two} we demonstrate that
the entropic uncertainty theorem takes the following form.
\begin{theorem}
  \label{thm:entropy-bounds}
If $\vec{a}$ is any eigenvector of a Fourier transform star graph, the
entropy of $\vec{a}$ satisfies
\begin{equation}
  \label{eq:94}
  \Ent(\vec{a}) \geq \frac12\log B + \log 2
\quad\text{and}\quad
  \REnt_{\sigma/(1-\sigma)}(\vec{a}) + \REnt_{-\sigma/(1+\sigma)}(\vec{a})
\geq \log B+2\log 2,
\end{equation}
for any $0\leq \sigma\leq 1$.
If $\vec{a}$ is any eigenvector of an equi-transmitting star graph then
\begin{equation}
  \label{eq:95}
  \Ent(\vec{a}) \geq \frac12\log(B-1) + \log 2
\quad\text{and}\quad
  \REnt_{\sigma/(1-\sigma)}(\vec{a}) + \REnt_{-\sigma/(1+\sigma)}(\vec{a})
\geq \log (B-1)+2\log 2.
\end{equation}
\end{theorem}

We find it notable that the bound \eqref{eq:94} is the average of
the maximum and minimum possible entropies for graph eigenvectors.

In our results on star graphs in section \ref{sec:five} and section
\ref{sec:equi-transmit} we establish the lower-bounds and prove the
existence of limiting eigenvectors that essentially achieve the
bounds, i.e.\ maximally scarred states.

\begin{theorem}
  \label{thm:fourier_star_main}
Consider a star graph with $B$ bonds, a Fourier transform scattering matrix
at the central vertex, and incommensurate bond lengths. 
Then there
exists a subsequence $(k_{n_j})\subseteq(k_n)$ such that the corresponding
normalised eigenvectors $\vec{a}_{n_j}$ satisfy
\begin{equation}
  \label{eq:114a}
  \vec{a}_{n_j} \to \vec{a}^{\mathcal{F}}
\end{equation}
as $j\to \infty$, where  $\vec{a}^{\mathcal{F}}$ is a vector with
Shannon entropy
\begin{equation}
  \label{eq:116a}
  \Ent(\vec{a}^{\mathcal F}) =\frac12\log B + 2\log2 + \Ord\!\left(
\frac{\log B}{B^{1/2}}\right),
\end{equation}
and the R\'enyi entropy is
\begin{equation}
  \label{eq:141a}
  \REnt_\infty(\vec{a}^{\mathcal{F}}) + \REnt_{-1/2}(\vec{a}^{\mathcal{F}}) =
\ln B + 2\log2 + \Ord(B^{-1/2}).
\end{equation}
\end{theorem}

For equi-transmitting star graphs we prove the following result in
section \ref{sec:equi-transmit}.
\begin{theorem}
  \label{thm:equi-transmitting_main}
Consider a star graph with $B$ bonds, an equi-transmitting scattering
matrix at the central vertex, and incommensurate bond lengths.
Then there is a subsequence $(k_{n_j})\subseteq (k_n)$ such that
the corresponding normalised eigenvectors $\vec{a}_{n_j}$ satisfy
  \begin{equation}
    \label{eq:infinity_minus_2}
\vec{a}_{n_j} \to \vec{a}^E
  \end{equation}
as $j\to\infty$, where
\begin{equation}
  \label{eq:133a}
 \Ent(\vec{a}^E) =  \frac12 \log(B-1) + 2\log 2,
\end{equation}
and
\begin{equation}
  \label{eq:144a}
  \REnt_\infty(\vec{a}^E) + \REnt_{-1/2}(\vec{a}^E)
= \log(B-1) + 2\log2 + \Ord(B^{-1/2}).
\end{equation}
\end{theorem}

The Shannon entropy \eqref{eq:Entropy} was studied for quantum graphs
in \cite{kam:eof}.  The authors used the entropic uncertainty
principle to prove lower bounds for the entropy of eigenfunctions for
regular graphs with equi-transmitting boundary conditions, and for
equi-transmitting star graphs.  They also calculate the average value
of the entropy of star graphs with a Kirchhoff scattering matrix.
Several of the results of \cite{kam:eof} were repeated for the R\'enyi
entropy in \cite{ras:RaT} as well as for a different generalisation of
the Shannon entropy (the Tsallis entropy) that we do not consider
here.

To compare the entropy bounds for existing eigenfunction localisations
in the literature, the limiting eigenvectors $\vec{a}^{\mathrm{N}}$ of
the star graph with Kirchhoff scattering matrices in \cite{ber:nqe}
are localised equally on two bonds of the graph, and zero mass
elsewhere, so the entropies are
\begin{equation}
\label{eq:26}
\Ent(\vec{a}^{\mathrm N}) = \REnt_\rho(\vec{a}^{\mathrm N}) = 2\log 2,
\end{equation}
while the entropic uncertainty principle furnishes the lower bound
\begin{equation}
\label{eq:27}
\Ent(\vec{a}) \geq \log2 + \Ord(B^{-1})
\end{equation}
and
\begin{equation}
\label{eq:28}
  \REnt_{\sigma/(1-\sigma)}(\vec{a}) + \REnt_{-\sigma/(1+\sigma)}(\vec{a})
\geq 2\log2 +\Ord(B^{-1})
\end{equation}
for eigenvectors of Kirchhoff star graphs.

The bounds in theorem \ref{thm:entropy-bounds}, and theorem 
\ref{thm:fourier_star_main} are proved in section \ref{sec:five}.
The vectors $\vec{a}^\curlyF$ in \eqref{eq:114a} have approximately
half the mass concentrated on the \emph{first} bond of the graph,
and the remaining mass equidistributed.  In section \ref{sec:new_four}
we add further explanation about the process behind theorem
\ref{thm:fourier_star_main}, and explain how to construct eigenfunctions
(half) localised on other bonds of the graph, and calculate their
entropies.  Some of the calculations from sections \ref{sec:five}
and \ref{sec:new_four} have been placed in appendices.

\section{Star graphs with non-Kirchhoff scattering matrices}
\label{sec:five}

For a star graph with $B$ bonds, scattering matrix $\Sigma$ at the
central vertex and Kirchhoff conditions at the external vertices, the
$2B\times 2B$ bond scattering matrix is (see appendix \ref{app:a})
\begin{equation}
\label{eq:64}
  S \coloneq \left( \begin{array}{cc}
           0 & \Sigma \\
           I & 0 
         \end{array} \right).
\end{equation}
We are going to consider the two cases of $\Sigma$ being
a $B\times B$ Fourier transform matrix, and a $B\times B$ equi-transmitting
matrix.  In contrast to the well-studied case where the central vertex
has Kirchhoff boundary conditions, the Fourier transform and equi-transmitting
cases are expected to produce generic spectral statistics in the limit
$B\to\infty$.

\subsection{Sketch of the main ideas}
\label{sec:three_point_one}
The $B\times B$ Fourier transform matrix $\mathcal{F}_B$ is defined to
be the matrix with $jk$th entry equal to
\begin{equation}
  \label{eq:84}
  \frac{1}{\sqrt{B}}\rme^{2\pi\rmi(j-1)(k-1)/B}.
\end{equation}
The matrix $\mathcal{F}_B$ is unitary.

Equi-transmitting matrices were introduced in \cite{har:qgw} as scattering
matrices of quantum graphs.  They are very closely related to
\emph{complex conference matrices} studied in the combinatorics literature.
An $B\times B$ complex matrix $E$ is equi-transmitting if it is unitary with 
diagonal entries $0$ and all other entries having equal absolute value.

As multiplication by a row or column by a pure phase does not change
the equi-transmitting property, we will assume that the equi-transmitting
matrices used are of the form
\begin{equation}
  \label{eq:45}
  E_B = \frac1{\sqrt{B-1}} \left( \begin{array}{ccccc}
                                  0 & 1 & 1 & \cdots & 1 \\
                                  1 & 0 & * & \cdots & * \\
                                  1 & * & 0 & \cdots & * \\
                                  \vdots & \vdots & \vdots & \ddots & \vdots \\
                                  1 & * & * & \cdots & 0 
                                \end{array} \right),
\end{equation}
however our method would work with only minor changes for an arbitrary
equi-trasmitting matrix, provided that its first row is identical to the
transpose of the first column.

We define three vectors belonging to $\C^B$:
\begin{equation}
  \label{eq:299}
  \vec{e} \coloneq \frac1{\sqrt{B}}\left( \begin{array}{c}
               1 \\
               1 \\
               \vdots \\
               1 
             \end{array} \right), \qquad
  \vec{u}_1 \coloneq \left( \begin{array}{c}
               1 \\
               0 \\
               \vdots \\
               0 
             \end{array} \right), \qquad
  \tilde{\vec{u}}_1 \coloneq \frac1{\sqrt{B-1}}\left( \begin{array}{c}
               0 \\
               1 \\
               \vdots \\
               1 
             \end{array} \right).
\end{equation}
It is relatively easy to see that
\begin{equation}
  \label{eq:46}
  {\mathcal{F}}_B \vec{e} = \vec{u}_1 \qquad\text{and}\qquad 
{\mathcal{F}}_B \vec{u}_1 = \vec{e}.
\end{equation}
It therefore follows that the vectors
\begin{equation}
  \label{eq:47}
  \vec{e} \pm \vec{u}_1
\end{equation}
are eigenvectors of $\mathcal{F}_B$ with eigenvalue $\pm1$, which moreover
are strongly enhanced in the first component by a factor roughly
$\sqrt{B}$. The Shannon entropy of $\vec{e}\pm\vec{u}_1$ is
\begin{equation}
\label{eq:25}
\frac12\log B + \log 2 \mp \frac1{B^{1/2}}\log(B^{1/2}\pm1).
\end{equation}
Starting from this point we may construct eigenvectors of the
scattering matrix of a $B$-bond Fourier star graph that are strongly
localised on the first bond by a factor about $\sqrt{B}$.

The chief difficulty will be the fact that the eigenvectors
$\vec{e}\pm \vec{u}_1$ belong to degenerate eigenspaces.  Indeed the
matrix ${\mathcal{F}}_B$ has eigenvalues $\pm1, \pm\rmi$ with the
dimensions of the corresponding eigenspaces approximately $B/4$ for
each eigenvalue \cite{aus:icw, mcc:eae}.
This degeneracy means that we cannot immediately conclude that a 
subsequence of eigenfunctions converges to the localised state.

By a careful choice of bond lengths, we are able to resolve this 
degeneracy, and indeed construct highly-localised eigenvectors.

The case of equi-transmitting scattering matrices at the centre of
the star graph proceeds similarly.  In this case, the pertinent
results (which may be easily checked) are
\begin{equation}
  \label{eq:50}
  E_B \vec{u}_1 = \tilde{\vec{u}}_1\qquad\text{and}\qquad 
E_B \tilde{\vec{u}}_1 = \vec{u}_1,
\end{equation}
so that the localised eigenvectors of $E_B$ are
\begin{equation}
  \label{eq:63}
  \vec{u}_1 \pm \tilde{\vec{u}}_1.
\end{equation}
In fact, because $\vec{u}_1$ and $\tilde{\vec{u}}_1$ are orthogonal,
the results for equi-transmitting matrices lead to comparatively cleaner
formul\ae\ (compare \eqref{eq:113} with \eqref{eq:128} below,
for example).

\subsection{Some spectral facts}
\label{sec:three_point_two}

The fact that we deal with $2B\times 2B$ scattering matrices of the form 
\eqref{eq:64} rather than the unitary $B\times B$ matrices ${\mathcal{F}}_B$
or $E_B$ themselves will not be a major technical hurdle.  The
following standard result allows us  to relate eigenvectors of the
central scattering matrix with the full quantum evolution operator matrix.
\begin{lemma}
  \label{lem:double}
Suppose that $\Pi$ and $\Sigma$ are $B\times B$ matrices and that
 $\vec{u}\in\C^B$ is an eigenvector of $\Sigma\Pi$ with eigenvalue $\nu\neq 0$. 
Then the vectors
\begin{equation}
  \label{eq:71}
  \left( \begin{array}{c}
           \pm\nu^{1/2}\vec{u} \\
           \Pi\vec{u} 
         \end{array} \right) \in\C^{2B},
\end{equation}
where $\nu^{1/2}$ is some square root of $\nu$, are eigenvectors of the
$2B\times 2B$ matrix
\begin{equation}
  \label{eq:85}
T\coloneq  \left( \begin{array}{cc}
           0 & \Sigma \\
           \Pi & 0 
         \end{array} \right)
\end{equation}
with eigenvalues $\pm\nu^{1/2}$.  
Moreover, all eigenvectors of $T$ with eigenvalue different from $0$
are of the form \eqref{eq:71}.
\end{lemma}
We may remark that if $\nu\neq0$ the vectors \eqref{eq:71} are 
linearly independent.

\dimostrazione Clearly the vectors \eqref{eq:71} are non-zero, so we
proceed by direct calculation:
\begin{align}
  \left( \begin{array}{cc}
           0 & \Sigma \\
           \Pi & 0 
         \end{array} \right) \left( \begin{array}{c}
                                      \pm\nu^{1/2}\vec{u} \\
                                      \Pi\vec{u} 
                                    \end{array} \right) &=
\left( \begin{array}{c}
         \Sigma\Pi\vec{u} \\
         \pm\nu^{1/2}\Pi\vec{u} 
       \end{array} \right) \nonumber \\
&= \left( \begin{array}{c}
            \nu \vec{u} \\
            \pm\nu^{1/2}\Pi\vec{u} 
          \end{array} \right) \nonumber \\
&= \pm\nu^{1/2}\left( \begin{array}{c}
            \pm\nu^{1/2}\vec{u} \\
            \Pi\vec{u} 
          \end{array} \right),
  \label{eq:86}
\end{align}
proving the first assertion. Contrariwise, suppose that $\vec{v}=\left(
\begin{array}{c} \vec{x} \\ \vec{y} \end{array}\right)$ is an eigenvector
of $T$ with eigenvalue $\lambda\neq0$.  From $T\vec{v} =
\lambda\vec{v}$ and $T^2\vec{v} = \lambda^2\vec{v}$, we conclude that 
\begin{equation}
  \label{eq:146}
  \vec{y} = \frac1{\lambda} \Pi\vec{x}
  \quad \mbox{and} \quad
  \Sigma\Pi\vec{x} = \lambda^2\vec{x},
\end{equation}
so $\vec{x}$ is an eigenvector of $\Sigma\Pi$ and
\begin{equation}
  \label{eq:148}
  \vec{v} = \frac1\lambda \left(
\begin{array}{c} \lambda\vec{x} \\ \Pi\vec{x} \end{array}\right),
\end{equation}
which is of the form \eqref{eq:71} up to an (irrelevant) scalar multiple.
\finire

We want to find a minimum theoretical bound for the entropy of 
eigenfunctions of star graphs with Fourier transform and equi-transmitting
scattering matrices.  A standard tool to accomplish this is the \emph{Entropic
Uncertainty Principle}, conjectured by Kraus \cite{kra:coa} and proved
by Maassen and Uffink \cite{maa:geu}.
Recall that two notions of the entropy of $\vec{a}$, were defined in
\eqref{eq:Entropy} and \eqref{eq:135} as
\begin{equation}
  \Ent(\vec{a}) \coloneq -\sum_{b=1}^{2B} \frac{|a_b|^2}{\|\vec{a}\|^2} \ln
\left( \frac{|a_b|^2}{\|\vec{a}\|^2} \right);\qquad
  \REnt_\rho(\vec{a}) \coloneq -\frac1\rho\ln\Bigg( \sum_{b=1}^{2B}
  \left( \frac{|a_b|^2}{\|\vec{a}\|^2} \right)^{1+\rho}\Bigg).
\end{equation}

\begin{theorem}(\cite{maa:geu})
  \label{thm:entropic}
Let $U=(U_{ij})$ be a unitary matrix.  For any complex vector $\vec{v}$,
\begin{equation}
\label{eq:138}
  \Ent(\vec{v}) + \Ent(U\vec{v}) \geq -\log \left( \max_{i,j}
|U_{ij}|^2 \right),
\end{equation}
and for any $0\leq\sigma\leq1$,
\begin{equation}
  \label{eq:137}
  \REnt_{\sigma/(1-\sigma)}(\vec{v}) + \REnt_{-\sigma/(1+\sigma)}(U\vec{v})
\geq -\log \left( \max_{i,j} |U_{ij}|^2 \right).
\end{equation}
If $\vec{v}$ is an eigenvector of $U$, then
\begin{equation}
  \label{eq:91}
  \Ent(\vec{v}) \geq -\frac12 \log \left( \max_{i,j}
|U_{ij}|^2 \right)
\end{equation}
and
\begin{equation}
  \label{eq:238}
  \REnt_{\sigma/(1-\sigma)}(\vec{v}) + \REnt_{-\sigma/(1+\sigma)}(\vec{v})
\geq -\log \left( \max_{i,j} |U_{ij}|^2 \right).
\end{equation}
\end{theorem}

We remark that \eqref{eq:138} and \eqref{eq:91} may be obtained as the
limit $\sigma\to0$ of \eqref{eq:137} and \eqref{eq:238} respectively.

For a star graph the quantum evolution operator is of the form
\begin{equation}
  \label{eq:87}
  U = \left( \begin{array}{cc}
               0 & \rme^{\rmi k L}\Sigma \\
               \rme^{\rmi k L} & 0 
             \end{array} \right), \qquad 
 \rme^{\rmi k L}\coloneq \diag\{\rme^{\rmi k L_1},\ldots,\rme^{\rmi k L_B}\},
\end{equation}
so it will not help to apply theorem \ref{thm:entropic} to $U$ \emph{directly}
as the entries of $U$ of absolute value $1$ will produce a zero in the
right-hand side of \eqref{eq:91}.  We could apply the theorem to the matrix
$U^2$ to get a non-trivial result; however we can do better.
We note from the preceding discussion that eigenvectors of $U$ are of
the form
\begin{equation}
  \label{eq:88}
  \vec{a} = \left( \begin{array}{c}
                     \pm\nu^{1/2}\vec{x} \\
                     \rme^{\rmi k L}\vec{x} 
                   \end{array} \right)
\end{equation}
where $\vec{x}$ is an eigenvector of the unitary matrix 
$\rme^{\rmi k L}\Sigma \rme^{\rmi k L}$, with eigenvalue $\nu\in\mathbb{T}$.

We define entropies of $\vec{x}\in\C^B$ by
\begin{equation}
  \label{eq:entropy}
  \ent(\vec{x}) \coloneq - \sum_{j=1}^B \frac{|x_j|^2}{\|\vec{x}\|^2} \log\left(
\frac{|x_j|^2}{\|\vec{x}\|^2} \right) \quad\text{and}\quad
  \rent_\rho(\vec{x}) \coloneq - \frac1\rho\ln\Bigg( \sum_{j=1}^{B}
  \left( \frac{|x_j|^2}{\|\vec{x}\|^2} \right)^{1+\rho}\Bigg),
\end{equation}
Since the absolute value of the $j$th entry of $\vec{a}$ of \eqref{eq:88}
is the same as the $(j+B)$th entry, and (for that very reason)
\begin{equation}
  \label{eq:90}
  \| \vec{a} \|^2 = 2\|\vec{x}\|^2,
\end{equation}
we have
\begin{equation}
  \REnt_\rho(\vec{a})
  = -\frac1\rho\ln\Bigg( 2\sum_{j=1}^{B}
  \left( \frac{|x_j|^2}{2\|\vec{x}\|^2} \right)^{1+\rho}\Bigg)
  = \rent_\rho(\vec{x}) + \log 2.
\label{eq:REnt-rent}
\end{equation}
and by taking the limit $\rho\to0$, or by direct calculation,
\begin{equation}
  \Ent(\vec{a}) = \ent(\vec{x}) + \log 2.
\label{eq:Ent-ent}
\end{equation}
A related equality was noted in \cite{kam:eof} (c.f.\ their Lemma 3).

\dimostrazionea{theorem \ref{thm:entropy-bounds}}
We apply theorem \ref{thm:entropic} to the vectors $\vec{x}$ and the
entropy \eqref{eq:entropy}.  Since vector $\vec{x}$ is an eigenvector
of $\rme^{\rmi k L}\Sigma\rme^{\rmi k L}$, and $\Sigma$ has bounded entries
in the cases in which we are interested, we get
\begin{equation}
  \label{eq:92}
  \ent(\vec{x}) \geq \frac12 \log B
\quad\text{and}\quad
  \rent_{\sigma/(1-\sigma)}(\vec{v}) + \rent_{-\sigma/(1+\sigma)}(\vec{v})
\geq \log B,
\end{equation}
if $\Sigma$ is a $B\times B$ Fourier transform matrix and
\begin{equation}
  \label{eq:93}
  \ent(\vec{x}) \geq \frac12 \log(B-1)
\quad\text{and}\quad
  \rent_{\sigma/(1-\sigma)}(\vec{v}) + \rent_{-\sigma/(1+\sigma)}(\vec{v})
\geq \log (B-1),
\end{equation}
if $\Sigma$ is a $B\times B$ equi-transmitting matrix.  These
lead \textit{via} \eqref{eq:Ent-ent} and \eqref{eq:REnt-rent} to entropy bounds
\eqref{eq:94} and \eqref{eq:95} respectively. \finire

A bound similar to \eqref{eq:95} for equi-transmitting star graphs was
derived by this method in \cite{kam:eof}.  In the following sections
we construct eigenfunctions that meet these bounds to leading order.

\subsection{Fourier scattering matrices}
\label{sec:three_point_three}

We consider matrices of the form
\begin{equation}
\label{eq:U_F}
  U_{\mathcal F} \coloneq U_{\mathcal F}(\kappa) \coloneq \left( 
\begin{array}{cc}
  0 & P(\kappa)\mathcal{F}_B \\
  P(\kappa) & 0 
\end{array} \right),
\end{equation}
where $P(\kappa)\coloneq\diag\{ \rme^{\rmi\kappa},1,\ldots,1 \}$.  If we 
choose $k_0$ and $L_0$ so that $P(\kappa) = D(k_0,L_0)$, which can always
be done, then \eqref{eq:U_F}
is the quantum evolution operator of a Fourier transform star graph with
bond lengths $L_0$.

By lemma \ref{lem:double} eigenvectors of $U_{\mathcal{F}}$ are of the form
\begin{equation}
  \label{eq:76}
  \vec{a} = \left( \begin{array}{c}
                     \pm\lambda^{1/2}\vec{x} \\
                     P(\kappa)\vec{x} 
                   \end{array} \right),
\end{equation}
where $\vec{x}$ is an eigenvector of the matrix $P(\kappa) \mathcal{F}_B
P(\kappa)$ with eigenvalue $\lambda\in{\mathbb{T}}$.  

In a calculation that is deferred to appendix \ref{app:b}, we show that there
are \emph{simple} 
eigenvectors $\vec{x}_\pm$ of $P(\kappa){\mathcal{F}}_B P(\kappa)$ with
the form
\begin{equation}
  \label{eq:666}
  \vec{x}_{\pm} = x_2\left( \begin{array}{c}
                            \cos\kappa\pm(B-\sin^2\kappa)^{1/2} \\
                            \vec{1}
                          \end{array}\right)  \in \C^{B}
\end{equation}
with $\vec{1}=(1,\ldots,1)^T\in\C^{B-1}$ and
$x_2\in\C\setminus\{0\}$.  It is further calculated in appendix
\ref{app:b} that
\begin{equation}
  \label{eq:8}
  \| \vec{x}_{\pm} \|^2 = 
2(B-\sin^2\kappa)^{1/2}((B-\sin^2\kappa)^{1/2}\pm\cos\kappa)|x_2|^2.
\end{equation}

We observe from \eqref{eq:666} that the absolute value squared of the
first component of $\vec{x}_\pm$ is 
\begin{equation}
  \label{eq:70}
 (\cos\kappa\pm(B-\sin^2\kappa)^{1/2})^2|x_2|^2=
\left( B \pm 2\sqrt{B}\cos\kappa + \cos2\kappa + \Ord(B^{-1/2})\right)|x_2|^2,
\end{equation}
and the remaining $B-1$ components have absolute value squared $|x_2|^2$.

Using \eqref{eq:76} and lemma \ref{lem:double} we can construct four
eigenvectors of $U_{\mathcal{F}}$.  We have now collected all the
ingredients required to prove our main result for this section:
\begin{proposition}
  \label{prop:fourier_star}
Consider a star graph with $B$ bonds, a Fourier transform scattering matrix
at the central vertex, and incommensurate bond lengths. 
Define
\begin{align}
  \label{eq:113}
  \vec{a}^{\mathcal F} &\coloneq 
\vec{a}^{\mathcal F}(\kappa,\epsilon_1,\epsilon_2)\\
&\coloneq \frac1{2D(\kappa)^{1/2}} \left(
\begin{array}{c}
  \epsilon_1\epsilon_2 
 \rme^{\pi\rmi/4+\rmi\kappa/2+\epsilon_2\rmi\Phi(\kappa)/2} 
  N(\kappa,\epsilon_2)^{1/2}\\
\epsilon_1 \rme^{\pi\rmi/4+\rmi\kappa/2+\epsilon_2\rmi\Phi(\kappa)/2} 
  N(\kappa,\epsilon_2)^{-1/2}\vec{1}\\
\epsilon_2\rme^{\rmi\kappa} N(\kappa,\epsilon_2)^{1/2}\\
  N(\kappa,\epsilon_2)^{-1/2}\vec{1}
\end{array}
\right),
\nonumber
\end{align}
for any $0<\kappa<\pi/2$, $\epsilon_1,\epsilon_2\in\{\pm 1\}$, where
\begin{equation}
\label{eq:34}
N(\kappa,\epsilon_2) \coloneq (B-\sin^2\kappa)^{1/2}+\epsilon_2\cos\kappa
\qquad\text{and}\qquad
D(\kappa)\coloneq (B-\sin^2\kappa)^{1/2},
\end{equation}
and $\Phi(\kappa)$ is a function whose value is given by \eqref{eq:61}.
Then there is a subsequence $(k_{n_j})\subseteq(k_n)$ such that the
corresponding normalised eigenvectors $\vec{a}_{n_j}$ satisfy
\begin{equation}
  \label{eq:114}
  \vec{a}_{n_j} \to \vec{a}^{\mathcal{F}}
\end{equation}
as $j\to \infty$.
\end{proposition}
Before giving the proof of proposition \ref{prop:fourier_star}, we consider
the properties of the limiting eigenvector $\vec{a}^{\mathcal{F}}$.
This will allow us to prove theorem \ref{thm:fourier_star_main}.

\dimostrazionea{theorem \ref{thm:fourier_star_main}}
We observe that $\vec{a}^{\mathcal F}$ is a superposition of a vector
with equal amplitude across the entire graph, and a component purely
localised on the first bond of the graph.  From \eqref{eq:Ent-ent},
the Shannon entropy of $\vec{a}^{\mathcal F}$ is given by
\begin{equation}
  \label{eq:115}
  \Ent(\vec{a}^{\mathcal F}) = \ent(\vec{x}_{\pm}) + \log 2,
\end{equation}
so we need to calculate $\ent(\vec{x}_\pm)$.

From the definition \eqref{eq:entropy}, and \eqref{eq:8} and \eqref{eq:70},
we can write
\begin{equation}
  \label{eq:72}
  \ent({\vec{x}_\pm}) = - \frac{N}{2D}\log\left( \frac{N}{2D} \right)
- (B-1)\frac{1}{2ND}\log\left(\frac1{2ND}\right),
\end{equation}
where $N$ and $D$ are abbreviations for $N(\kappa,\pm1)$ and $D(\kappa)$ 
introduced in \eqref{eq:34}.
We write \eqref{eq:72} in the equivalent form
\begin{equation}
  \label{eq:74}
  \ent(\vec{x}_\pm) = -\frac1{2D}\left( \left( N-\frac{B-1}N \right)\log N
- \left( N + \frac{B-1}N \right) \log 2D \right)
\end{equation}
and given that
\begin{equation}
  \label{eq:130}
  \frac{N}{2D}+\frac{B-1}{2ND} = 1,
\end{equation}
we have
\begin{equation}
  \label{eq:134}
  \ent(\vec{x}_\pm) = \log 2D - \frac{N^2-(B-1)}{N^2+(B-1)}\log N.
\end{equation}
With definitions \eqref{eq:34}, \eqref{eq:134} gives
\begin{equation}
\label{eq:22}
\ent(\vec{x}_\pm) = \log 2(B-\sin^2\kappa)^{1/2} \mp 
\frac{N\cos\kappa}{B-1\pm N\cos\kappa} \log\left(
(B-\sin^2\kappa)^{1/2}\pm\cos\kappa\right),
\end{equation}
where we have used 
\begin{align}
N^2 &= B-\sin^2\kappa \pm 2\cos\kappa(B-\sin^2\kappa)^{1/2}+\cos^2\kappa 
\nonumber \\
&= B-1 \pm 2(B-\sin^2\kappa)^{1/2}\cos\kappa + 2\cos^2\kappa \nonumber \\
&= B-1 \pm 2N\cos\kappa,
  \label{eq:75}
\end{align}
to simplify.
If $B$ is large, then,
\begin{equation}
  \label{eq:83}
  \ent(\vec{x}_\pm) = 
\frac12\log B + \log2 +\Ord\!\left(\frac{\log B}{B^{1/2}}\right)
\end{equation}
This means that for large values of $B$,
\begin{equation}
  \label{eq:116}
  \Ent(\vec{a}^{\mathcal F}) =\frac12\log B + 2\log2 + \Ord\!\left(
\frac{\log B}{B^{1/2}}\right)
\end{equation}
essentially (to leading order) meeting the bound \eqref{eq:94}.
This proves \eqref{eq:116a}.  

By considering the R\'enyi entropy, we may improve the result further,
essentially removing the $\log2$ term in \eqref{eq:83}.
With the same notation as above, for $0\leq\sigma\leq1$,
\begin{align}
  \rent_{\sigma/(1-\sigma)}(\vec{x}_\pm) + \rent_{-\sigma/(1+\sigma)}
  &(\vec{x}_\pm) \nonumber
  \\
  &= \frac{\sigma-1}{\sigma}\log \left(
    \left( \frac{N}{2D} \right)^{1/(1-\sigma)} + (B-1)\left( 
    \frac{1}{2ND} \right)^{1/(1-\sigma)}\right) \nonumber \\
  &\qquad + \frac{\sigma+1}{\sigma}\log \left( \left( \frac{N}{2D} 
    \right)^{1/(1+\sigma)} + (B-1) \left( \frac1{2ND} \right)^{1/(1+\sigma)}
    \right)\nonumber \\
  &= \frac{\sigma-1}{\sigma}\log 
    \left( N^{1/(1-\sigma)} + (B-1)N^{-1/(1-\sigma)}\right) \nonumber \\
  &\quad + \frac{\sigma+1}{\sigma}\log \left( N^{1/(1+\sigma)} + 
    (B-1) N^{-1/(1+\sigma)} \right).
    \label{eq:80}
\end{align}
Now we specialise to $\sigma\to1$.  Equation \eqref{eq:80} yields
\begin{equation}
  \label{eq:82}
  \rent_\infty(\vec{x}_\pm) + \rent_{-1/2}(\vec{x}_\pm) = -\log N +
  2\log \left( N^{1/2} + (B-1) N^{-1/2} \right).
\end{equation}
Inserting $N=((B-\sin^2\kappa)^{1/2}\pm\cos\kappa)$ we get
\begin{align}
  \label{eq:139}
  \rent_\infty(\vec{x}_\pm) + \rent_{-1/2}(\vec{x}_\pm) &=
2\ln(B-1)-2\ln((B-\sin^2\kappa)^{1/2}\pm\cos\kappa) \nonumber \\
&\qquad +2\ln \left( 1+ \frac{(B-\sin^2\kappa)^{1/2}\pm\cos\kappa}{B-1} \right) 
\nonumber\\
&= \ln B + \Ord(B^{-1/2}),
\end{align}
for $B$ large.  Thus the bound given by \eqref{eq:92}:
\begin{align}
  \label{eq:140}
  \rent_\infty(\vec{x}_\pm) + \rent_{-1/2}(\vec{x}_\pm) \geq \log B,
\end{align}
is achieved up to terms negligible in the size of the graph.  The 
R\'enyi entropy of $\vec{a}^{\mathcal{F}}$ satisfies
\begin{equation}
  \label{eq:141}
  \REnt_\infty(\vec{a}^{\mathcal{F}}) + \REnt_{-1/2}(\vec{a}^{\mathcal{F}}) =
\ln B + 2\log2 + \Ord(B^{-1/2}),
\end{equation}
as $B\to\infty$, which is \eqref{eq:141a}.\finire

\dimostrazionea{proposition \ref{prop:fourier_star}}
We have
\begin{equation}
  \label{eq:117}
  \vec{a}^{\mathcal{F}}(\kappa,\epsilon_1,\epsilon_2) =
 \frac1{\sqrt{2}} \left( \begin{array}{c}
         \epsilon_1 \lambda^{1/2} \hat{\vec{x}}_{\pm} \\
         P(\kappa)\hat{\vec{x}}_\pm
       \end{array} \right),
\end{equation}
with $\lambda$ given by \eqref{eq:62}, and the sign $\pm$ chosen
according to $\epsilon_2=\pm1$, and $\hat{\vec{x}}_{\pm}$ are
normalised versions of $\vec{x}_\pm$ from \eqref{eq:66}.  

If we choose $k_0, L$ such that $k_0L_1=\kappa$ and $k_0L_j=1$
for $j=2,\ldots,B$, then $\rme^{\rmi k_0L}=P(\kappa)$, and 
$U_{\mathcal{F}}$ is the quantum evolution operator $D(k_0,L)S$. It
follows from lemma \ref{lem:double} that $\vec{a}^{\mathcal F}$ 
is an eigenvector of $U_{\mathcal{F}}$ with eigenvalue $\epsilon_1
\lambda^{1/2}$.  Because $\hat{\vec{x}}_{\pm}$ are simple eigenvectors,
it further follows that $\vec{a}^{\mathcal F}$ is a simple eigenvector
of the star graph.

The required convergence then follows from theorem \ref{thm:cdv}.
\finire

\subsection{Eigenvectors localised on other bonds}
\label{sec:three_point_four}
Proposition \ref{prop:fourier_star} proves the existence of a limiting 
eigenvector of a Fourier star graph with enhanced mass concentrated on
the first bond of the graph.  We can modify the construction to prove
the existence of limiting eigenvectors with enhanced mass on a different
bond as we now describe.

Let $P_j(\kappa)$ be the diagonal $B\times B$ matrix with $j$th
diagonal entry equal to $\rme^{\rmi\kappa}$ and all other entries
equal to $1$.  This generalises the matrix $P(\kappa)=P_1(\kappa)$
introduced in \eqref{eq:U_F}.  The na\"\i ve modification of
\eqref{eq:U_F} with $P_j(\kappa)$ instead of $P(\kappa)$ does not
produce an eigenvector half-localized on the bond $j$; instead, the
result is an eigenvector with half of the mass spread throughout
the graph and the other half split equally among the bonds $j$ and
$B-j+2$.  This is described in more detail in section~\ref{sec:new_four}.

To produce an eigenvector half-localizing on the bond $j$ alone, we
will need some extra steps.  Let $j\in\{1,\ldots,B\}$ and define the
$B\times B$ diagonal matrix
\begin{equation}
\label{eq:20}
R_j = \diag\{ \rme^{-2\pi\rmi(j-1)(m-1)/B + \pi\rmi(j-1)^2/B} : m=1,\ldots,B\}.
\end{equation}
The $mn$th entry of the matrix $R_j\curlyF_BR_j$ is
\begin{align}
  &\frac1{\sqrt{B}} \rme^{-2\pi\rmi(j-1)(m-1)/B + \pi\rmi(j-1)^2/B}
\rme^{2\pi\rmi (m-1)(n-1)/B}
\rme^{-2\pi\rmi(j-1)(n-1)/B + \pi\rmi(j-1)^2/B} \nonumber \\
&=\frac1{\sqrt{B}} \rme^{2\pi\rmi(n-j)(m-j)/B}.
\end{align}
In other words, $R_j\curlyF_BR_j$ is a cyclically-permuted version of
$\curlyF_B$ with the first row and column moved to the $j$th position.
Then we have that $P_j(\kappa)R_j \curlyF_B
R_jP_j(\kappa)$ is a permuted version of $P(\kappa)\curlyF_B
P(\kappa)$.  

By permuting all vectors appearing in the argumentation of section
\ref{sec:three_point_three}, we find eigenvectors
$\tilde{\vec{x}}_\pm$ with simple eigenvalue $\tilde{\lambda}_\pm$,
entry squared \eqref{eq:70} in the $j$th component, and $|x_2|^2$ in
all other components.  Because entropy is invariant with respect to
permutations, $\ent(\tilde{\vec{x}}_\pm)= \ent(\vec{x}_\pm)$ as
calculated in \eqref{eq:22}, and similarly for
$\rent_\rho(\tilde{\vec{x}}_\pm)$.

By lemma \ref{lem:double}, there are eigenvectors of the matrix
\begin{equation}
\label{eq:23}
 \left( \begin{array}{cc}
          0 & P_j(\kappa)R_j\curlyF_B \\
          P_j(\kappa)R_j & 0 
        \end{array} \right)
\end{equation}
of the form 
\begin{equation}
\label{eq:24}
\tilde{\vec{a}} = \left( \begin{array}{c}
                   \pm\tilde{\lambda}^{1/2}_\pm\tilde{\vec{x}}_\pm \\
                   P_j(\kappa)R_j\tilde{\vec{x}}_\pm
                 \end{array} \right),
\end{equation}
and the same entropies as $\vec{a}^\curlyF$, and moreover have enhanced
amplitude on the $j$th bond.  These can be shown to be limiting
eigenvectors of the Fourier transform star graph by choosing
$k_0, L_0$ such that $D(k_0,L_0)=P_j(\kappa)R_j$ as was done in the
proof of proposition \ref{prop:fourier_star}.

\subsection{Sharpness of the Entropic Uncertainty Principle}
\label{sec:three_point_five}
Below the leading order behaviour, there is a difference of $\log2$
between the Shannon entropy bound arising out of the Entropic
Uncertainty Principle and the limiting eigenvectors
$\vec{a}^\curlyF$---compare \eqref{eq:94} with \eqref{eq:116a}.  The
same is true in fact for the Shannon entropy of the limiting
eigenvectors on equi-transmitting star graphs (equation
\eqref{eq:133a}) and Kirchhoff star graphs (equation \eqref{eq:26}
against the lower-bound \eqref{eq:27}).  This discrepancy could lead
to speculation that the Entropic Uncertainty Principle is not sharp.
It being prudent to investigate this possibility further we undertook
a numerical investigation of the simplest non-trivial case involving
the $3\times 3$ Fourier transform matrix $\curlyF_3$.  This has the
advantages that the matrices are small enough to allow a reasonable
exploration of the parameter space, as well as avoiding degeneracies
as $\curlyF_3$ has a simple spectrum.  A basis of eigenvectors of
$\curlyF_3$ is
\begin{equation}
\label{eq:29}
\left( \begin{array}{c} 1+\sqrt3 \\ 1 \\ 1 \end{array} \right),\quad
\left( \begin{array}{c} 1-\sqrt3 \\ 1 \\ 1 \end{array} \right),\quad
\left( \begin{array}{c} 0 \\ 1 \\ -1 \end{array} \right).
\end{equation}
The first listed vector in \eqref{eq:29} is $\vec{e}+\vec{u}_1$ as
described in section \ref{sec:three_point_one} and its entropy is
(from \eqref{eq:25})
\begin{equation}
\label{eq:30}
\ent(\vec{e}+\vec{u}_1)=
\frac12\log3 + \log 2 - \frac1{\sqrt{3}}\log(\sqrt3+1) \approx 0.662.
\end{equation}
Our findings may be summarised as follows: the calculations indicate
that the the Entropic Uncertainty bound as proved by Maassen and
Uffink \cite{maa:geu} \emph{is} sharp in this case; nevertheless our
eigenvector results cannot be improved in the sense that there are no
Fourier-transform-like matrices with lower entropy eigenvectors.

\begin{figure}
  \centering
     \includegraphics{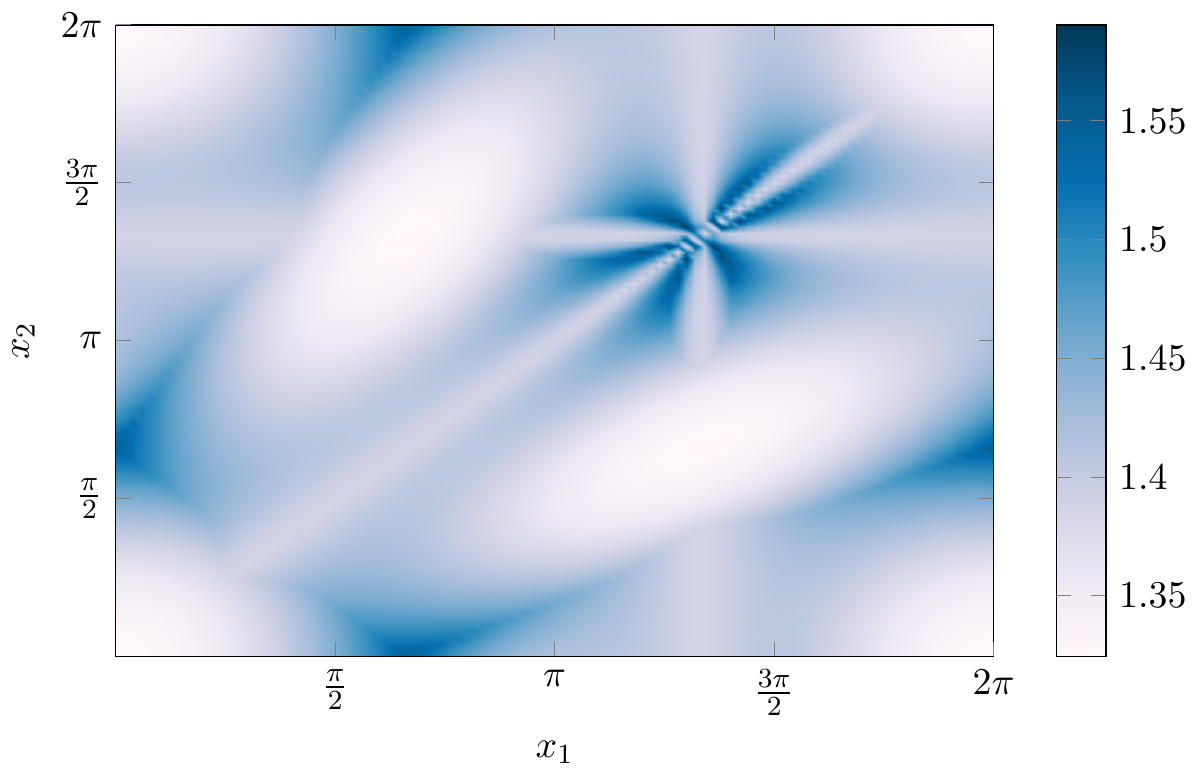}
    \caption{A plot of double the Shannon entropy of the eigenvector with 
least entropy for 
the matrices $\Upsilon(x_1,x_2)\curlyF_3$ where $\Upsilon(x_1,x_2)=\diag\{1,
\rme^{\rmi x_1}, \rme^{\rmi x_2}\}$.  Minimal entropy found is approximately
$2\ent(\vec{e}+\vec{u}_1)\approx 1.324$ from \eqref{eq:25}.  The
curious formation at $(\frac{4\pi}3,\frac{4\pi}3)$ is due to a
degeneracy in the spectrum of $\Upsilon\curlyF_3$ at this point.}
  \label{fig:zwei}
\end{figure}

The sharpness of the first part of Theorem~\ref{thm:entropic} when $U
= \curlyF_N$ is illustrated by taking $\vec{v} = \vec{e}_1 =
(1,0,0,\ldots)^T$ and, correspondingly, $U\vec{v} =
(1,1,1,\ldots)^T/\sqrt{N}$.  In this case $\REnt_{\rho}(\vec{v}) = 0$
and
\begin{equation}
  \label{eq:REnt_calculation_uniform}
  \REnt_\rho(U\vec{v}) = -\frac1\rho \log\left(
    \sum_{j=1}^N N^{-1-\rho} \right) = \log N,
\end{equation}
independently of $\rho$.  The value of the right-hand side of
inequalities \eqref{eq:138}--\eqref{eq:137} is also $\log N$.
 
Of course, the vector $\vec{v}$ chosen above is far from being an
eigenvector of $U$ and it is unsurprising that inequalities
\eqref{eq:91}--\eqref{eq:238} are not sharp.
In order to consider if there are Fourier-transform-like matrices with
eigenvectors with low entropy we consider the family of
 matrices $\Upsilon(x_1,x_2)\curlyF_3$ where
\begin{equation}
\label{eq:31}
\Upsilon(x_1,x_2)\coloneq \left( \begin{array}{ccc}
                               1 & 0 & 0 \\
                               0 & \rme^{\rmi x_1} & 0 \\
                               0 & 0 & \rme^{\rmi x_2} 
                             \end{array} \right).
\end{equation}
This family essentially parameterises the space of $3\times 3$ unitary
matrices with all entries having equal amplitude (such matrices are
called \emph{complex Hadamard matrices}).  We plot in figure 
\ref{fig:zwei} the Shannon entropy\footnote{For computational reasons, the
\emph{double} of the Shannon entropy is actually plotted.}
of the eigenvector with the lowest entropy
as $x_1$, $x_2$ vary.  The minimal entropy we find matches numerically
the value \eqref{eq:30}, at $(x_1,x_2)=(0,0)$, corresponding to
$\mathcal{F}_3$ itself, and at $(\frac{2\pi}3,\frac{4\pi}3)$ and
$(\frac{4\pi}3,\frac{2\pi}3)$ corresponding to cyclic permutations
of $\curlyF_3$ that were discussed in section \ref{sec:three_point_four}.
We conclude that there are not any complex Hadamard matrices with
eigenvectors with entropy closer to the bound of $\log 3$ than
the Fourier transform matrix and its permutations.

\section{Interpretation and generalisation}
\label{sec:new_four}

In the previous section we saw that by varying the length of
the first bond of the Fourier transform star graph we arrive at 
quantum limits strongly enhanced by a factor approximately $\sqrt{B}$ on
the first bond, with all other components having equal amplitude. To
produce localised eigenvectors on other bonds we need to vary all
bond lengths as described in section \ref{sec:three_point_four}.

In this section we explain what happens if we vary the length of
a single bond different from the first bond.  What we find is
an eigenstate that is strongly localised on two bonds of the
graph: the $j$th bond, the one that is varied, and the $(B+2-j)$th bond.
Approximately half the mass is shared equally between those two bonds,
and the remaining mass distributed on the remaining bonds.

The analysis turns out to be more complicated, but we are able as a
result to give an explanation for the especially simple form
\eqref{eq:666} of perturbed eigenvectors in the case where the first
bond length is varied, as well as supplying an alternative proof for
why the perturbed eigenvectors are simple.  Most of the technical
detail is in Appendix \ref{app:c}.

Let $P_j(\kappa)$ be the diagonal matrix $B\times B$ matrix with $j$th
diagonal entry equal to $\rme^{\rmi\kappa}$ and all other entries
equal to $1$ as above. Then take
\begin{equation}
\label{eq:UU_F}
  U_{\mathcal F} \coloneq U_{\mathcal F}(\kappa) \coloneq \left( 
\begin{array}{cc}
  0 & P_j(\kappa)\mathcal{F}_B \\
  P_j(\kappa) & 0 
\end{array} \right)
\end{equation}
in place of \eqref{eq:U_F}.  By lemma \ref{lem:double}, eigenvectors
of this $U_\curlyF$ are related to the eigenvectors of 
$P_j(\kappa)\curlyF_B P_j(\kappa)$ through formula \eqref{eq:71} as
we describe below.

Let $F_j(\kappa)\coloneq P_j(\kappa)^2\curlyF_B$.  If $\vec{f}$ is an
eigenvector of $F_j(\kappa)$ then $\vec{x}=P_j(\kappa)^{-1}\vec{f}$ is an
eigenvector of $P_j(\kappa)\curlyF_B P_j(\kappa)$ with the same
eigenvalue, and moreover
\begin{equation}
  \label{ne:16}
  \ent(\vec{x}) = \ent(\vec{f}),
\end{equation}
since the entries of $\vec{x}$ and $\vec{f}$ are different only by phases.

To an eigenvector $\vec{f}$ of $F_j(\kappa)$ with eigenvalue $\lambda$,
 lemma \ref{lem:double} furnishes eigenvectors
\begin{equation}
  \label{eq:9}
  \vec{a} = \left( \begin{array}{c}
                     \pm\lambda^{1/2}P_j(-\kappa)\vec{f} \\ \vec{f} 
                   \end{array} \right)
\end{equation}
of $U_\curlyF$, and due to \eqref{eq:Ent-ent} and \eqref{ne:16}, the
Shannon entropy of $\vec{a}$ is
\begin{equation}
  \label{eq:10}
  \Ent(\vec{a}) = \ent(\vec{f}) + \log 2.
\end{equation}

The matrix $F_j(\kappa)$ is more beneficial to study because it can be
written as a rank-$1$ perturbation of $\curlyF_B$:
\begin{equation}
  \label{ne:17}
  F_j(\kappa) = \left(I + (\rme^{2\rmi\kappa}-1) \vec{e}_j 
    \vec{e}_j^\dag \right)\curlyF_B,
\end{equation}
with $\vec{e}_j$ as the $j$th standard $\C^{B}$ basis vector.  In
particular, the matrix $F_j(\kappa)$ coincides with $\curlyF_B$ on a
subspace of co-dimension 1 --- the orthogonal complement to the vector
$\vec{e}_j$.

\begin{remark}
  \label{rem:rank_one}
  Eigenvectors and eigenvalues of rank-$1$ perturbations can be
  expressed in terms of those of the unperturbed matrix \cite{alb:smi,
    bog:mot} (See also \cite{cdv:plI}).  Usually these calculations
  are carried out for Hermitian matrices, but the case here with
  unitary matrices can be done easily analogously, as was done in
  \cite{bog:ssoII}.  The following fact is especially useful to us:
  if $\rme^{\rmi\phi}$ is
  an eigenvalue of $\curlyF_B$ for which the eigenspace
  $\Eig(\curlyF_B,\rme^{\rmi\phi})$ contains a vector $\vec{u}$ with
  $\vec{e}_j^\dag\vec{u}=0$, then $\rme^{\rmi\phi}$ is an eigenvalue
  of $F_j(\kappa)$ with multiplicity
  $\dim\Eig(\curlyF_B,\rme^{\rmi\phi}) - 1$.  We remark that this is
  valid for all $\kappa$ and also valid in the case when
  $\dim\Eig(\curlyF_B,\rme^{\rmi\phi}) = 1$, in which case
  $\rme^{\rmi\phi}$ is \emph{not} an eigenvalue of $F_j(\kappa)$.
\end{remark}

It follows that we need to understand the eigenvalues of the Fourier
matrix $\curlyF_B$ (see \cite{mcc:eae} for more complete information).
Explicit computation shows that $\curlyF_B^2$ is the matrix of the
permutation $\left(\begin{smallmatrix}
                   1 & 2 & 3 & \cdots & B-1 & B \\
                   1 & B & B-1 & \cdots & 3 & 2 
                   \end{smallmatrix}\right)$.  
This permutation leaves invariant
the element $1$ and, if $B$ is even, the element $B/2+1$.  Since it is
an involution, we conclude that $\curlyF_B^4 = I$ and the spectrum of
$\curlyF_B$ consists of the numbers $\pm1$ and $\pm \rmi$.

\begin{figure}
  \centering
  \begin{tikzpicture}
    [blob/.style={circle, radius=1mm, fill=white, draw=black, inner sep=1pt}]
    \draw[->] (-3,0) -- (3,0) node[anchor=north] {$\Re\lambda$};
    \draw[->] (0,-2.5) -- (0,2.5) node[anchor=south west] {$\Im\lambda$};
    \draw[thick]  (0,0) circle (2);
    \node at (0,-3) {(a)};
    \draw --plot[mark=x] coordinates {(2,0) (0,2) (-2,0) (0,-2)};
    \node[blob] at (30:2) {};
    \node[blob] at (220:2) {};
    \draw[->,>=latex] (2.2,0) arc (0:28:2.2);
    \draw[->,>=latex] (-2.2,0) arc (180:218:2.2);
    \begin{scope}[xshift=7cm]
      \draw[->] (-3,0) -- (3,0) node[anchor=north] {$\Re\lambda$};
      \draw[->] (0,-2.5) -- (0,2.5) node[anchor=south west] {$\Im\lambda$};
      \draw[thick]  (0,0) circle (2);
      \node at (0,-3) {(b)};      
      \draw --plot[mark=x] coordinates {(2,0) (0,2) (-2,0) (0,-2)};
      \node[blob] at (20:2) {};
      \node[blob] at (210:2) {};
      \node[blob] at (100:2) {};
      \node[blob] at (315:2) {};
      \draw[->,>=latex] (2.2,0) arc (0:18:2.2);
      \draw[->,>=latex] (-2.2,0) arc (180:208:2.2);
      \draw[->,>=latex] (0,2.2) arc (90:98:2.2);
      \draw[->,>=latex] (0,-2.2) arc (270:313:2.2);
    \end{scope}
  \end{tikzpicture}
  \caption{Schematic depiction of eigenvalues of the matrix
    $F_j(\kappa)$ as a function of $\kappa$ when $j=1$ and when $j
    \neq 1$.}
  \label{fig:eig_movement}
\end{figure}
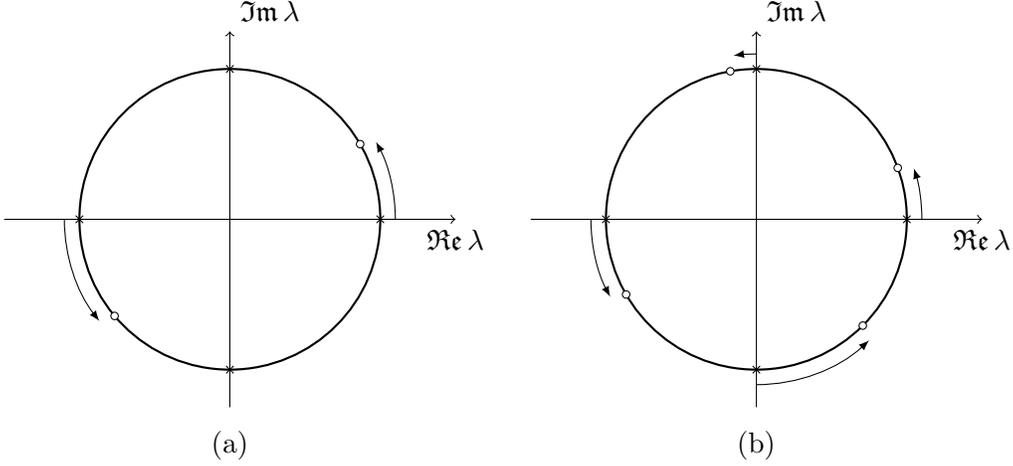

We start with the case $j=1$.  Let $\vec{e} = (1,1,1,\ldots)^T$ and
$\vec{u}_1=\vec{e}_1$.  As noted in \eqref{eq:47}, the vectors
$\vec{e}\pm \vec{u}_1$ are eigenvectors of $\curlyF_B$ with
eigenvalues $\pm1$.  Complete these two vectors to an orthonormal
basis
$\{\vec{e}+\vec{u}_1, \vec{e}-\vec{u}_1, \vec{w}_3, \ldots,
\vec{w}_B\}$
of $\curlyF_B$, which we can always do because $\curlyF_B$ is unitary.
Since $\vec{e}_1 = \vec{u}_1$ is a linear combination of the first two
vectors, each of the eigenvectors $\vec{w}_3,\ldots, \vec{w}_B$ is orthogonal to
$\vec{e}_1$ and therefore still an eigenvector of $F_1(\kappa)$ with
the same eigenvalue.  The two other eigenvalues of $F_1(\kappa)$ must
be different from $\pm1$ by Remark~\ref{rem:rank_one} but must remain
close to $\pm1$ for small $\kappa$ (by classical perturbation theory
results).  This is shown schematically in figure~\ref{fig:eig_movement}(a).  
The corresponding
eigenvectors must remain orthogonal to $\vec{w}_3,\ldots, \vec{w}_B$
and are therefore linear combinations of $\vec{e} \pm \vec{u}_1$,
i.e.\ can be written in the general form,
\begin{equation}
  \label{eq:77a}
  \vec{x} = \left( \begin{array}{c}
                     x_1 \\
                     x_2 \vec{1} 
                   \end{array} \right), \quad x_1, x_2\in\C.
\end{equation}
This explains the form of the eigenvectors in equation \eqref{eq:666}.
We get qualitatively similar results as in section
\ref{sec:three_point_three} if $B$ is even and $j=B/2+1$.

Now take general $j$ different from 1 and (if $B$ is even) from
$B/2+1$.  Suppose for simplicity that $B\geq5$.  Let
$\vec{u}_j \in\C^B$ be the vector with $1$s in the $j$th and
$(B+2-j)$th entries, and zeros in all other positions, and
$\vec{v}_j\in\C^B$ be the vector with a $1$ in $j$th position, $-1$ in
the $(B+2-j)$th position, and zeros in all other entries. For example,
in the case $j=2$ we have
\begin{equation}
  \vec{u}_2 = \left( \begin{array}{c}
 0 \\ 1 \\ 0 \\ \vdots \\ 0 \\ 1 
 \end{array} \right)\qquad\text{and}\qquad
  \vec{v}_2 = \left( \begin{array}{c}
 0 \\ 1 \\ 0 \\ \vdots \\ 0 \\ -1 
 \end{array} \right).
\end{equation}
Then define
\begin{equation}
\label{ne:2}
  \begin{aligned}
    \wplus &\coloneq \curlyF_B \vec{u}_j + \vec{u}_j \\
    \wminus &\coloneq \curlyF_B \vec{u}_j - \vec{u}_j \\
    \wplusi &\coloneq \rmi\curlyF_B \vec{v}_j - \vec{v}_j\\
    \wminusi &\coloneq \rmi\curlyF_B \vec{v}_j + \vec{v}_j.
  \end{aligned}
\end{equation}
If $B\geq5$, these vectors are non-zero.  Noting that
$\curlyF_B^2 \vec{u}_j = \vec{u}_j$ and $\curlyF_B^2 \vec{v}_j = -\vec{v}_j$ 
(by the permutation properties of $\curlyF_B^2$), we see that
$\wplus, \wminus, \wplusi,\wminusi$ are eigenvectors of $\curlyF_B$
with eigenvalues respectively: $+1, -1, +\rmi, -\rmi$.  Proceed as
before: complete this set to find an orthonormal basis
\begin{equation}
  \label{eq:basis_j}
  \{\wplus, \wminus, \wplusi, \wminusi, \vec{w}_5, \ldots \vec{w}_B\}.  
\end{equation}
Since
\begin{equation}
  \label{eq:14}
  \vec{e}_j = \frac14 (\wplus - \wminus - \wplusi + \wminusi),
\end{equation}
the eigenvectors $\vec{w}_5,\ldots \vec{w}_B$ are orthogonal to
$\vec{e}_j$ and thus remain eigenvectors of the perturbed matrix
$F_j(\kappa)$.  The other eigenvalues of $F_j(\kappa)$ must be
different from $\pm1$ and $\pm\rmi$ by Remark~\ref{rem:rank_one}.  In
fact, they move with $\kappa$ as indicated in figure 
\ref{fig:eig_movement}(b).  The corresponding eigenvectors are linear
combinations of $\{\wplus, \wminus, \wplusi, \wminusi\}$, and therefore 
can be written in the form
\begin{equation}
  \label{ne:5a}
  \vec{f} = \beta_1 \curlyF_B\vec{u}_j + \beta_2 \rmi\curlyF_B\vec{v}_j +
  \beta_3 \vec{u}_j + \beta_4 \vec{v}_j,
\end{equation}
A calculation presented in appendix \ref{app:c} results in
\begin{equation}
  \label{ne:15a}
  \begin{aligned}
    \beta_1 &= Z\rme^{-\rmi\phi}\cot\phi \\
    \beta_2 &= Z\rme^{-\rmi\phi} \\
    \beta_3 &= Z\cot\phi \\
    \beta_4 &= \rmi Z,
  \end{aligned}
\end{equation}
where $Z\in\C\setminus\{0\}$, and $\rme^{\rmi\phi}$ is the perturbed 
eigenvalue, which is one of the solutions $\lambda$ to 
\begin{equation}
  \label{eq:19}
  \lambda^4 - \frac{\lambda^3 (\rme^{2\rmi\kappa}-1)}{\sqrt{B}}
\rme^{2\pi\rmi(j-1)^2/B}
- \frac{\lambda (\rme^{2\rmi\kappa}-1)}{\sqrt{B}}\rme^{-2\pi\rmi (j-1)^2/B} 
- \rme^{2\rmi\kappa} = 0.
\end{equation}
The norm of $\vec{f}$ is
\begin{equation}
  \label{eq:20a}
   \|\vec{f}\|^2 =
 |Z|^2 \cosec^2\phi \left( 4 + \frac8{\sqrt{B}}\left( \cos^3\phi
 \cos\left(\frac{2\pi(j-1)^2}B \right) + \sin^3\phi
  \sin \left( \frac{2\pi(j-1)^2}B \right) \right) \right)
\end{equation}
and the entropy satisfies
\begin{equation}
  \label{ne:113a}
  \ent(\vec{f}) = \frac12\log B + 2\log 2 - \frac12
+ \Ord\left( \frac{\log B}{B^{1/2}} \right).
\end{equation}
The corresponding entropy of the vector $\vec{a}$ given by \eqref{eq:9} is,
from \eqref{eq:10},
\begin{equation}
  \label{eq:21}
  \Ent(\vec{a}) = \frac12\log B + 3\log 2 - \frac12
+ \Ord\left( \frac{\log B}{B^{1/2}} \right).
\end{equation}
We observe that this is further from the theoretical bound
\eqref{eq:94} than we could achieve by varying the first bond, as
reported in section \ref{sec:three_point_three}.  The perturbed
eigenfunctions now have approximately half the mass concentrated on
the $j$th and the $(B+2-j)$th bond, and the remaining components are
not equal, but vary from bond-to-bond.  This is the explanation for
why the entropy \eqref{eq:21} is further from the bound; the
difference is a fundamental consequence of the shape of the
eigenfunctions and cannot be removed by switching to a different
entropy.

\section{Equi-transmitting boundary conditions}
\label{sec:equi-transmit}
We now turn our attention to quantum star graphs with an equi-transmitting
scattering matrix at the centre.  To that effect we consider the
$2B\times 2B$ unitary matrices of the
form
\begin{equation}
  \label{eq:111}
  U_E \coloneq U_E(\kappa) \coloneq \left( \begin{array}{cc}
                               0 & P(\kappa)E_B \\
                               P(\kappa) & 0 
                             \end{array} \right)
\end{equation}
where $E_B$ is an equi-transmitting matrix of the form \eqref{eq:45}.

\begin{proposition}
  \label{prop:equi-transmitting}
Consider a star graph with $B$ bonds, an equi-transmitting scattering
matrix $E_B$ of the form \eqref{eq:45} at the central vertex, and
incommensurate bond lengths. Let
  \begin{equation}
    \label{eq:128}
    \vec{a}^E \coloneq \vec{a}^E(\kappa,\epsilon_1,\epsilon_2) \coloneq
\frac1{2\sqrt{B-1}} \left( \begin{array}{c}
 \epsilon_1\epsilon_2 \rme^{\pi\rmi(1-\epsilon_2)/4+\rmi\kappa/2} \sqrt{B-1} \\
 \epsilon_1 \rme^{\pi\rmi(1-\epsilon_2)/4+\rmi\kappa/2}\vec{1} \\
                              \epsilon_2 \rme^{\rmi\kappa}\sqrt{B-1} \\
                              \vec{1} 
                            \end{array} \right)
  \end{equation}
for any $\epsilon_1,\epsilon_2\in\{\pm1\}$ and $0<\kappa\leq\pi$ such that
$\epsilon_2\rme^{\rmi\kappa}$ is \emph{not} an eigenvalue of $E_B$.
Then there is a subsequence $(k_{n_j})\subseteq (k_n)$ such that
the corresponding normalised eigenvectors $\vec{a}_{n_j}$ satisfy
  \begin{equation}
    \label{eq:infinity_minus_1}
\vec{a}_{n_j} \to \vec{a}^E
  \end{equation}
as $j\to\infty$.
\end{proposition}

\dimostrazione
Eigenvectors of $U_E$ are, by lemma \ref{lem:double}, of the
form
\begin{equation}
  \label{eq:118}
  \vec{a} = \left( \begin{array}{c}
                     \pm\mu^{1/2}\vec{y} \\
                     P(\kappa)\vec{y} 
                   \end{array} \right)
\end{equation}
where $\vec{y}$ is an eigenvector of $P(\kappa)E_BP(\kappa)$ with
eigenvalue $\mu\in\T$.

If we suppose
\begin{equation}
  \label{eq:119}
  \vec{y}  = \left( \begin{array}{c} y_1 \\
                      y_2\vec{1} \end{array} \right)
= y_1 \vec{u}_1 + \sqrt{B-1} y_2 \tilde{\vec{u}}_1
\end{equation}
using definitions \eqref{eq:299}.  We have
\begin{equation}
  \label{eq:120}
  P(\kappa)\vec{y}  = \left( \begin{array}{c} \rme^{\rmi \kappa}y_1 \\
                      y_2\vec{1} \end{array} \right)
= \rme^{\rmi\kappa} y_1 \vec{u}_1 + \sqrt{B-1} y_2 \tilde{\vec{u}}_1,
\end{equation}
and by \eqref{eq:50},
\begin{align}
  E_B P(\kappa)\vec{y}  
&= \rme^{\rmi\kappa} y_1 \tilde{\vec{u}}_1 + \sqrt{B-1} y_2 \vec{u}_1
\nonumber \\
&= \left( \begin{array}{c} y_2\sqrt{B-1} \\
           \frac1{\sqrt{B-1}}\rme^{\rmi\kappa} y_1\vec{1} \end{array} \right).
  \label{eq:121}
\end{align}
Finally,
\begin{equation}
  \label{eq:122}
 P(\kappa)  E_B P(\kappa)\vec{y}  
= \left( \begin{array}{c} \rme^{\rmi\kappa} y_2\sqrt{B-1} \\
           \frac1{\sqrt{B-1}}\rme^{\rmi\kappa} y_1\vec{1} \end{array} \right).
\end{equation}
The eigenequation $P(\kappa)E_BP(\kappa)\vec{y}=\mu\vec{y}$ is soluble
in non-zero $\vec{y}$ if and only if $\mu$ satisfies
\begin{equation}
  \label{eq:123}
  \begin{aligned}
    \rme^{\rmi \kappa}y_2 \sqrt{B-1} &= \mu y_1, \\
    \frac1{\sqrt{B-1}}\rme^{\rmi \kappa} y_1 &= \mu y_2.
  \end{aligned}
\end{equation}
The system \eqref{eq:123} requires $\mu^2=\rme^{2\rmi\kappa}$, whence
\begin{equation}
  \label{eq:124}
  \mu = \pm \rme^{\rmi\kappa}
\end{equation}
and
\begin{equation}
  \label{eq:125}
  y_1 = \pm\sqrt{B-1} y_2.
\end{equation}
Define
\begin{equation}
  \label{eq:126}
  \vec{y}_\pm \coloneq y_2 \left( \begin{array}{c} \pm\sqrt{B-1} \\
                         \vec{1} \end{array} \right),
\end{equation}
and the normalised versions
\begin{equation}
  \label{eq:127}
  \hat{\vec{y}}_\pm \coloneq \frac1{\sqrt{2(B-1)}} \left( \begin{array}{c} 
  \pm\sqrt{B-1} \\ \vec{1} \end{array} \right).
\end{equation}
The vectors $\vec{y}_\pm$ (and $\hat{\vec{y}}_\pm$) are eigenvectors
with eigenvalue $\pm\rme^{\rmi\kappa}$.  Moreover, it follows
from the same argument as in section \ref{sec:new_four}
that all other eigenvectors of $P(\kappa)E_BP(\kappa)$ are eigenvectors
of $E_B$, so, provided that $\pm\rme^{\rmi\kappa}$ is not an
eigenvalue of $E_B$, $\vec{y}_\pm$ is simple.

The vector $\vec{a}^E$ is given by
\begin{equation}
  \label{eq:129}
  \vec{a}^E = \frac1{\sqrt2} \left( \begin{array}{c}
                                      \epsilon_1 \mu^{1/2}\hat{\vec{y}}_\pm \\
                                      P(\kappa)\hat{\vec{y}}_\pm 
                                    \end{array} \right)
\end{equation}
with the $\pm$ sign chosen according to $\epsilon_2=\pm1$.  Due to
the above it may be deduced that this is a simple eigenvector of $U_E$.
For choice of bond lengths and $k_0$ as in the proof
of proposition \ref{prop:fourier_star}, $\vec{a}^E$ is a simple eigenvector
of the quantum evolution operator, and hence we get the convergence
\eqref{eq:infinity_minus_1} due to theorem \ref{thm:cdv}. \finire

For certain values\footnote{Specifically if $B\equiv2\;\mathrm{mod}\;4$
with $B-1$ a prime; see Corollary 3.5 of \cite{har:qgw}.} of $B$
it is possible to find equi-transmitting matrices $E_B$ that
are symmetric.  For such matrices, $\pm1$ are the only eigenvalues
and hence any $\kappa$ with $0<\kappa<\pi$ can be taken in 
proposition \ref{prop:equi-transmitting}.  The matrix $E_6$ shown in 
\eqref{eq:89} is such an example.

\dimostrazionea{theorem \ref{thm:equi-transmitting_main}}
From \eqref{eq:Ent-ent} and \eqref{eq:129},
\begin{equation}
  \label{eq:131}
  \Ent(\vec{a}^E) = \ent(\hat{\vec{y}}_\pm) + \log 2,
\end{equation}
where $\ent(\hat{\vec{y}}_\pm)$ may be calculated from \eqref{eq:134}
except that now
\begin{equation}
  \label{eq:142}
  N \coloneq D\coloneq \sqrt{B-1}.
\end{equation}
We get
\begin{equation}
  \ent(\hat{\vec{y}}_\pm) = \frac12 \log(B-1) + \log 2.
  \label{eq:132}
\end{equation}
This leads to
\begin{equation}
  \label{eq:133}
 \Ent(\vec{a}^E) =  \frac12 \log(B-1) + 2\log 2,
\end{equation}
which is \eqref{eq:133a} and 
should be compared with the theoretical bound \eqref{eq:95}
for equi-transmitting star graph eigenvectors. Again, to leading
order, the bound has been reached.  However, we can again improve this
by considering the R\'enyi entropy.  If we consider \eqref{eq:82}
with $N, D$ as in \eqref{eq:142} for the case $\hat{\vec{y}}_\pm$,
we obtain
\begin{align}
  \rent_\infty(\hat{\vec{y}}_\pm) + \rent_{-1/2}(\hat{\vec{y}}_\pm)
&= -\frac12\log(B-1) + 2\log \left( (B-1)^{1/4} + (B-1)^{3/4} \right) 
\nonumber \\
&= \log(B-1) + 2\log \left( 1+ (B-1)^{-1/2} \right) 
  \label{eq:143}
\end{align}
for large values of $B$.  We have achieved the bound \eqref{eq:93}
up to terms negligible in $B$.  For the R\'enyi entropy of $\vec{a}^E$
we have
\begin{equation}
  \label{eq:144}
  \REnt_\infty(\vec{a}^E) + \REnt_{-1/2}(\vec{a}^E)
= \log(B-1)  +2\log 2+ 2\log \left( 1+ (B-1)^{-1/2} \right).
\end{equation}
which leads to \eqref{eq:144a}.\finire

These results mirror the results that we proved in section
\ref{sec:three_point_three} for the Fourier transform scattering
matrix, and varying the length of the first bond.  Because we do not
have complete information about the spectrum and eigenvectors of
general equi-transmitting matrices, we cannot repeat the analysis of
the vectors reached by varying other bond lengths than the first one.
Whether it is possible to do this in specific cases, such as when the
equi-transmitting matrices are constructed from Dirichlet characters
\cite{har:qgw} where $B-1$ is a prime number is an interesting
question that we leave for future study.

\subsubsection*{Acknowledgements}
The authors are grateful to 
\href{https://www.math.u-psud.fr/~nonnenma/index.html}{St\'ephane Nonnenmacher}
for a helpful discussion of the Entropic Uncertainty Principle.
GB acknowledges partial support from the NSF under grant DMS-1410657.


\appendix
\section{Bond-scattering matrix of a star graph} %
\label{app:a}
We suppose a star graph with $B$ bonds (and $B+1$ vertices) has at the
central vertex the scattering matrix $\Sigma$, and we suppose
Kirchhoff conditions at the leaves (exterior vertices). For vertices
of degree one, this condition reduces to simple reflection of
waves. The argument below can be modified easily to allow other conditions at
the leaves.

We order the directed bonds so that the first $B$ bonds point inwards
to the central vertex, and the subsequent bonds are the reversals.
The wave-function for $k>0$ on the $j$th bond can be written
\begin{equation}
  \psi_j(x) = \ain_j \rme^{-\rmi k x} + \aout_j \rme^{\rmi k x}.
\end{equation}
and because we will be assuming bond $B+j$ is the reversal of bond $j$, 
\begin{align}
  \psi_{B+j}(x) &= \psi_j(L_j-x) \nonumber \\
  &= \ain_j \rme^{-\rmi k L_j + \rmi k x} + \aout_j \rme^{\rmi k L_j - 
\rmi k x},
\end{align}
whence, for $1\leq j \leq B$,
\begin{equation}
  \label{eq:2}
  \ain_{j+B} = \rme^{\rmi k L_j} \aout_j \qquad\text{and}\qquad
  \aout_{j+B} = \rme^{-\rmi k L_j} \ain_j.
\end{equation}
The $B\times B$ matrix $\Sigma$ controls scattering at the central vertex 
so
\begin{align}
\Sigma   \left( \begin{array}{c}
           \ain_{B+1} \\
           \vdots \\
           \ain_{2B} 
         \end{array} \right) &=
 \left( \begin{array}{c}
           \aout_{B+1} \\
           \vdots \\
           \aout_{2B} 
         \end{array} \right) 
\nonumber \\
&=\left( \begin{array}{c}
           \rme^{-\rmi k L_1} \ain_{1} \\
           \vdots \\
           \rme^{-\rmi k L_B}\ain_{B} 
         \end{array} \right) \qquad\text{using \eqref{eq:2},}
\nonumber \\
&= \diag\{\rme^{-\rmi k L_1},
           \ldots, 
           \rme^{-\rmi k L_B}\} \left( \begin{array}{c}
           \ain_{1} \\
           \vdots \\
           \ain_{B} 
         \end{array} \right).
  \label{eq:3}
\end{align}
At the leaves of the star graph waves are reflected which imposes
\begin{align}
  \ain_j &= \aout_j,\qquad j=1,\ldots,B, \nonumber \\
  &= \rme^{-\rmi k L_j} \ain_{j+B},
\end{align}
using (\ref{eq:2}) once more.  We have
\begin{equation}
 \left( \begin{array}{c} \ain_1 \\ \vdots \\ \ain_B \end{array} \right)
=  \left( \begin{array}{c}
           \rme^{-\rmi k L_1} \ain_{B+1} \\
           \vdots \\
           \rme^{-\rmi k L_B}\ain_{2B} 
         \end{array} \right) 
\label{eq:4}
=  \diag\{\rme^{-\rmi k L_1},
           \ldots, 
           \rme^{-\rmi k L_B}\} \left( \begin{array}{c}
           \ain_{B+1} \\
           \vdots \\
           \ain_{2B} 
         \end{array} \right).
\end{equation}
We can combine \eqref{eq:3} and \eqref{eq:4} into a single system: denoting
\begin{equation}
  \label{eq:5}
  \vec{a} = (\ain_1, \ain_2, \ldots, \ain_{2B})^T,
\end{equation}
we have
\begin{equation}
  \label{eq:6}
  D(-k) \vec{a} = \left( \begin{array}{cc}
                           0 & \Sigma \\
                           I & 0 
                         \end{array} \right) \vec{a}.
\end{equation}
Equivalently, $(I-D(k)S)\vec{a}=0$, with
\begin{equation}
  \label{eq:7}
  S = \left( \begin{array}{cc}
                           0 & \Sigma \\
                           I & 0 
                         \end{array} \right),
\end{equation}
the bond scattering matrix.
\section{Perturbing the length of the first bond}
\label{app:b}
In this appendix we calculate eigenvectors $\vec{x}$ of the matrix
$P(\kappa)\curlyF_B P(\kappa)$ introduced in section
\ref{sec:three_point_three}.  Let us look for $\vec{x}$ of the form
\begin{equation}
  \label{eq:77}
  \vec{x} = \left( \begin{array}{c}
                     x_1 \\
                     x_2 \vec{1} 
                   \end{array} \right)\in\C^B
\end{equation}
where $x_1, x_2\in\C$ and $\vec{1}=(1,\ldots,1)^T\in\C^{B-1}$.  We have
\begin{equation}
  \label{eq:78}
  P(\kappa)\vec{x} = \left( \begin{array}{c}
                     \rme^{\rmi\kappa} x_1 \\
                     x_2 \vec{1} 
                   \end{array} \right).
\end{equation}
We can write
\begin{equation}
  \left( \begin{array}{c}
           \rme^{\rmi\kappa} x_1 \\
           \vec{1}x_2 
         \end{array} \right) = \sqrt{B} x_2 \vec{e} + (\rme^{\rmi\kappa}x_1-
x_2) \vec{u}_1,
\end{equation}
so that it is easy to see from \eqref{eq:46} that
\begin{equation}
  \label{eq:48}
\mathcal{F}_B P(\kappa) \vec{x} 
= \sqrt{B} x_2 \vec{u}_1 + (\rme^{\rmi\kappa}x_1 -x_2) \vec{e}.
\end{equation}
Thus, the first component of $P(\kappa)\mathcal{F}_B P(\kappa)\vec{x}$ is 
       \begin{equation}
         \label{eq:49}
         \rme^{\rmi\kappa}\left( \sqrt{B}x_2 + \frac{\rme^{\rmi\kappa}x_1-
x_2}{\sqrt{B}} \right),
       \end{equation}
and all remaining components take the same value $(\rme^{\rmi\kappa}x_1-
x_2)/\sqrt{B}$.  
The eigenvector equation
\begin{equation}
  \label{eq:51}
  P(\kappa) {\mathcal F}_B P(\kappa) \vec{x} = \lambda \vec{x},
\end{equation}
can be expressed using \eqref{eq:77} as the $2\times 2$ linear system
\begin{equation}
  \label{eq:52}
  \left( \begin{array}{cc}
           \rme^{2\rmi\kappa}/\sqrt{B} & \rme^{\rmi\kappa}(B-1)/\sqrt{B} \\
           \rme^{\rmi\kappa}/\sqrt{B} & -1/\sqrt{B} \\
         \end{array} \right)
\left( \begin{array}{c}
         x_1 \\
         x_2 
       \end{array} \right) = \lambda \left( \begin{array}{c}
         x_1 \\
         x_2 
       \end{array} \right).
\end{equation}
The characteristic equation of the matrix in \eqref{eq:52} is
\begin{equation}
  \label{eq:53}
  \det \left( \begin{array}{cc}
                \lambda - \frac{\rme^{2\rmi\kappa}}{\sqrt{B}} & 
-\rme^{\rmi\kappa}\frac{B-1}{\sqrt{B}} \\
                -\frac{\rme^{\rmi\kappa}}{\sqrt{B}} & \lambda+\frac1{\sqrt{B}}
              \end{array} \right) = 0,
\end{equation}
which can be expressed as
\begin{align}
                \left(\lambda - \frac{\rme^{2\rmi\kappa}}{\sqrt{B}}\right)
\left( \lambda+\frac1{\sqrt{B}} \right) -
\rme^{2\rmi\kappa}\frac{B-1}{B} &= 0 \nonumber\\
\Rightarrow\quad \lambda^2 - \frac{1}{\sqrt{B}}(\rme^{2\rmi\kappa}-1)\lambda
- \rme^{2\rmi\kappa} &= 0.
  \label{eq:54}
\end{align}
Now 
\begin{align}
  \rme^{2\rmi\kappa}-1 &= \rme^{\rmi\kappa}(\rme^{\rmi\kappa} - 
\rme^{-\rmi\kappa}) \nonumber \\
&= 2\rmi\rme^{\rmi\kappa}\sin \kappa,  \label{eq:112}
\end{align}
so we have
\begin{equation}
  \label{eq:57}
  (\rme^{2\rmi \kappa} - 1 )^2 = -4\sin^2\kappa \rme^{2\rmi\kappa},
\end{equation}
and (for substitution into the quadratic formula)
\begin{equation}
  \label{eq:58}
  \frac1B (\rme^{2\rmi\kappa}-1)^2 + 4 \rme^{2\rmi\kappa} = 4
 \left( 1-\frac1B\sin^2\kappa \right)\rme^{2\rmi\kappa}.
\end{equation}
The solutions $\lambda$ to \eqref{eq:54} satisfy
\begin{align}
  \lambda &= \frac{B^{-1/2}(\rme^{2\rmi \kappa}-1) \pm 2(1-B^{-1}\sin^2
\kappa)^{1/2}\rme^{\rmi\kappa}}{2} \nonumber \\
&= \frac{\rme^{2\rmi \kappa}-1 \pm 2(B-\sin^2
\kappa)^{1/2}\rme^{\rmi\kappa}}{2\sqrt{B}} \nonumber \\
&= \frac{2(\pm(B-\sin^2\kappa)^{1/2}+\rmi\sin\kappa)\rme^{\rmi\kappa}}
{2\sqrt{B}},
  \label{eq:59}
\end{align}
using \eqref{eq:112}.  Writing
\begin{equation}
  \label{eq:60}
  (\pm(B-\sin^2\kappa)^{1/2}+\rmi\sin\kappa) = \sqrt{B}\rme^{\pi\rmi/2\pm
\rmi\Phi(\kappa)}
\end{equation}
where
\begin{equation}
  \label{eq:61}
  \tan\Phi(\kappa) \coloneq -\frac{\sqrt{B-\sin^2\kappa}}{\sin\kappa},
\end{equation}
we get
\begin{equation}
  \label{eq:62}
  \lambda = \rme^{\pi\rmi/2+\rmi\kappa\pm\rmi\Phi(\kappa)}
\end{equation}
and neither of the two values of $\lambda$ are in the set $\{\pm1, \pm\rmi\}$,
if $0<\kappa<\pi/2$.

Let us now turn to the eigenvectors corresponding to the eigenvalues
$\lambda$.  From \eqref{eq:52}, we get
  \begin{equation}
    \label{eq:65}
\begin{aligned}
    \rme^{\rmi\kappa} B x_2 + \rme^{2\rmi\kappa}x_1 - \rme^{\rmi\kappa} x_2
 &= \lambda\sqrt{B}x_1 \\
\rme^{\rmi\kappa}x_1 - x_2 &= \lambda\sqrt{B}x_2,
\end{aligned}
  \end{equation}
which are equivalent to
\begin{equation}
  \label{eq:67}
  x_1 = \rme^{-\rmi\kappa}(\lambda\sqrt{B}+1)x_2.
\end{equation}
We note that from \eqref{eq:59},
\begin{align}
  \lambda\sqrt{B}+1 &= (\pm (B-\sin^2\kappa)^{1/2} + \rmi \sin\kappa
+\rme^{-\rmi\kappa})\rme^{\rmi\kappa} \\
&= (\cos\kappa \pm (B-\sin^2\kappa)^{1/2})\rme^{\rmi\kappa},
  \label{eq:68}
\end{align}
so \eqref{eq:67} becomes
\begin{equation}
  \label{eq:79}
  x_1 = (\pm(B-\sin^2\kappa)^{1/2}+\cos\kappa)x_2.
\end{equation}
The upshot is that we find eigenvectors of $P(\kappa){\mathcal F}_B
P(\kappa)$ of the form
\begin{equation}
  \label{eq:66}
  \vec{x}_{\pm} = x_2\left( \begin{array}{c}
                            \cos\kappa\pm(B-\sin^2\kappa)^{1/2} \\
                            \vec{1}
                          \end{array}\right)  \in \C^{B}
\end{equation}
where $x_2$ becomes a normalisation constant.
 The component
of $\vec{x}_\pm$ that contributes to the eigenfunction on the first bond of 
the star graph is its first component, and this is enhanced by a
factor approximately $B^{1/2}$ in magnitude above the components on all 
other bonds (which are in fact equal).

We now want to show that $\vec{x}_{\pm}$ are simple eigenvectors. 
This follows from the discussion in section~\ref{sec:new_four}.  To 
see it directly we may observe that in extending $\{\vec{x}_+, \vec{x}_-\}$
to an orthogonal basis $  \{ \vec{x}_+, \vec{x}_{-}, \vec{x}_3,\ldots ,
\vec{x}_B\}$, leads to the first component of $\vec{x}_j$, $j\geq3$ being
zero, and consequently
\begin{equation}
  \label{eq:99}
  P(\kappa)\vec{x}_j = \vec{x}_j,
\end{equation}
for any value of $\kappa$.  It then follows that 
\begin{equation}
  \label{eq:102}
  \mathcal{F}_B \vec{x}_j = \lambda_j \vec{x}_j,
\end{equation}
i.e.\ the vectors $\vec{x}_j$ are eigenvectors of $\mathcal{F}_B$, and
therefore $\lambda_j\in\{ \pm1, \pm\rmi\}$.  As we have already
observed the eigenvalues of $\vec{x}_{\pm}$ do not belong to the
aforementioned set, thus they are simple eigenvectors.

To assist with calculating the entropy of $\vec{x}_\pm$, we note that
\begin{align}
  \|\vec{x}_\pm\|^2 &= \left( (\cos\kappa \pm (B-\sin^2\kappa)^{1/2})^2
+ B-1 \right)|x_2|^2 \nonumber \\
                &= \left( B -\sin^2\kappa \pm 2\cos\kappa(B-
\sin^2\kappa)^{1/2} + \cos^2\kappa + B -1 \right)|x_2|^2 \nonumber \\
                &= \left( 2B \pm 2\cos\kappa(B-\sin^2\kappa)^{1/2}
-2\sin^2\kappa\right)|x_2| \nonumber \\
&= 2(B \pm \cos\kappa(B-\sin^2\kappa)^{1/2}-\sin^2\kappa)|x_2|^2 \nonumber \\
&= 2(B-\sin^2\kappa)^{1/2}((B-\sin^2\kappa)^{1/2}\pm\cos\kappa)|x_2|^2.
  \label{eq:69}
\end{align}
\section{Perturbation of other bond lengths}
\label{app:c}
In this appendix we calculate eigenvectors of the matrix $F_j(\kappa)$ 
described in \eqref{ne:17}. Recall that this is a rank-one perturbation
of the $B\times B$ Fourier transform matrix $\curlyF_B$.

We refer to certain quantities defined in section \ref{sec:new_four}:
the vectors $\vec{u}_j$ and $\vec{v}_j$, and the vectors
$\wplus, \wminus, \wplusi,\wminusi$ defined in \eqref{ne:2},
which are eigenvectors of $\curlyF_B$ with eigenvalues respectively: $+1, -1,
+\rmi, -\rmi$.  We repeat the important identity
\begin{equation}
\label{ne:3}
  \vec{e}_j = \frac14( \wplus - \wminus - \wplusi + \wminusi).
\end{equation}
where $\vec{e}_j$ is the $j$th standard basis vector.

These calculations properly hold only if $B\geq 5$.  If $B=3$ or $4$ then
some details change (the eigenspace $\Eig(\curlyF_B,-\rmi)$ is trivial
and the vector $\wminusi=0$).

\subsection{Eigenvectors of the perturbed matrix}
To save notation we write $K(\kappa) \coloneq \rme^{2\rmi\kappa}-1$.

We look for eigenvectors of $F_j(\kappa)$ of the form
\begin{equation}
  \label{ne:4}
  \vec{f} = \aplus\wplus + \aminus\wminus + \aplusi\wplusi + \aminusi
\wminusi,
\end{equation}
which can be re-expressed as
\begin{equation}
  \label{ne:5}
  \vec{f} = \beta_1 \curlyF_B\vec{u}_j + \beta_2 \rmi\curlyF_B\vec{v}_j +
\beta_3 \vec{u}_j + \beta_4 \vec{v}_j,
\end{equation}
with
\begin{equation}
  \label{ne:6}
  \left( \begin{array}{c}
           \beta_1 \\
           \beta_2 \\
           \beta_3 \\
           \beta_4 
         \end{array} \right)  = \left( \begin{array}{cccc}
                                         1 & 1 & 0 & 0 \\
                                         0 & 0 & 1 & 1 \\
                                         1 & -1 & 0 & 0 \\
                                         0 & 0 & -1 & 1 
                                       \end{array} \right) \left( 
\begin{array}{c}
  \aplus \\
  \aminus \\
  \aplusi \\
  \aminusi 
\end{array} \right).
\end{equation}

Since $F_j(\kappa) = (\curlyF_B + K(\kappa) \vec{e}_j\vec{e}_j^\dag \curlyF_B)$
we use the fact that $\vec{f}$ is a linear combination of eigenvectors of
$\curlyF_B$ to write
\begin{align}
  F_j(\kappa) \vec{f}  &= (I + K(\kappa)\vec{e}_j\vec{e}_j^\dag)\curlyF_B
\vec{f} \nonumber \\
&=(I + K(\kappa)\vec{e}_j\vec{e}_j^\dag) 
( \aplus\wplus - \aminus\wminus + \rmi\aplusi\wplusi -\rmi \aminusi\wminusi).
\end{align}
We then use (\ref{ne:3}), and orthogonality of the $\vec{w}_i$ to express
\begin{equation}
  \label{ne:7}
  F_j(\kappa) \vec{f} =
( \aplus\wplus - \aminus\wminus + \rmi\aplusi\wplusi -\rmi \aminusi\wminusi)
+\frac14 K(\kappa)A (\wplus - \wminus - \wplusi + \wminusi)
\end{equation}
where
\begin{equation}
  \label{ne:8}
  A=\frac14 \left( \aplus \|\wplus\|^2 + \aminus \|\wminus\|^2 -\rmi
\aplusi \| \wplusi \|^2 -\rmi \aminusi \| \wminusi\|^2 \right).
\end{equation}
Equating components in the eigenvalue equation
\begin{equation}
  \label{ne:9}
  F_j(\kappa) \vec{f} = \rme^{\rmi\phi}\vec{f},
\end{equation}
we find that the vector $\vecalpha=(\aplus, \aminus, \aplusi, \aminusi)^T$ 
satisfies the linear equation system
\begin{equation}
  \label{ne:10}
  \left({\renewcommand{\arraystretch}{1.2} \begin{array}{cccc}
 1 + \frac{K(\kappa)}{16}\|\wplus\|^2 & \frac{K(\kappa)}{16}\|\wminus\|^2 & 
 -\frac{\rmi K(\kappa)}{16}\|\wplusi\|^2 & 
 -\frac{\rmi K(\kappa)}{16}\|\wminusi\|^2 \\
 -\frac{K(\kappa)}{16}\|\wplus\|^2 & -1-\frac{K(\kappa)}{16}\|\wminus\|^2 & 
 \frac{\rmi K(\kappa)}{16}\|\wplusi\|^2 & 
 \frac{\rmi K(\kappa)}{16}\|\wminusi\|^2 \\
 -\frac{K(\kappa)}{16}\|\wplus\|^2 & -\frac{K(\kappa)}{16}\|\wminus\|^2 & 
 \rmi+\frac{\rmi K(\kappa)}{16}\|\wplusi\|^2 & 
 \frac{\rmi K(\kappa)}{16}\|\wminusi\|^2 \\
 \frac{K(\kappa)}{16}\|\wplus\|^2 & \frac{K(\kappa)}{16}\|\wminus\|^2 & 
 -\frac{\rmi K(\kappa)}{16}\|\wplusi\|^2 & 
 -\rmi -\frac{\rmi K(\kappa)}{16}\|\wminusi\|^2 
         \end{array}} \right) \vecalpha = \rme^{\rmi\phi}\vecalpha.
\end{equation}
From (\ref{ne:6}) it follows that
\begin{equation}
  \label{ne:11}
  \vecalpha = \frac12 \left( \begin{array}{cccc}
                               1 & 0 & 1 & 0 \\
                               1 & 0 & -1 & 0 \\
                               0 & 1 & 0 & -1 \\
                               0 & 1 & 0 & 1 
                             \end{array} \right)\vecbeta,
\end{equation}
with $\vecbeta=(\beta_1,\beta_2,\beta_3,\beta_4)^T$.  We may 
re-write \eqref{ne:10} as an equation for $\vecbeta$,
using (\ref{ne:6}) and (\ref{ne:11}), as
\begin{equation}
  \label{ne:12}
  \left( {\renewcommand{\arraystretch}{1.2} \begin{array}{cccc}
 0 & 0 & 1 & 0 \\
 0 & 0 & 0 & -\rmi \\
 1+\frac{K(\kappa)}{16}W_1 & -\frac{\rmi K(\kappa)}{16} W_2 & 
 \frac{K(\kappa)}{16}W_3 & \frac{\rmi K(\kappa)}{16} W_4 \\
 \frac{K(\kappa)}{16}W_1 & -\rmi-\frac{\rmi K(\kappa)}{16} W_2 & 
 \frac{K(\kappa)}{16}W_3 & \frac{\rmi K(\kappa)}{16} W_4
         \end{array}} \right)\vecbeta = \rme^{\rmi\phi}\vecbeta,
\end{equation}
with
\begin{equation}
  \label{ne:13}
  \begin{aligned}
    W_1 &\coloneq \|\wplus\|^2 + \|\wminus\|^2 \\
    W_2 &\coloneq \|\wplusi\|^2 + \|\wminusi\|^2\\
    W_3 &\coloneq \|\wplus\|^2 - \|\wminus\|^2 \\
    W_4 &\coloneq \|\wplusi\|^2 - \|\wminusi\|^2.
  \end{aligned}
\end{equation}
We observe immediately from \eqref{ne:12} that
\begin{equation}
  \label{ne:14}
  \begin{aligned}
    \beta_3 &= \rme^{\rmi\phi} \beta_1, \\
   -\rmi \beta_4 &= \rme^{\rmi\phi} \beta_2,
  \end{aligned}
\end{equation}
and, on subtracting the fourth row from the third, 
\begin{align}
\nonumber
  \beta_1 + \rmi\beta_2 &= \rme^{\rmi\phi} \beta_3 - \rme^{\rmi\phi} \beta_4\\
\Rightarrow\quad \rme^{-\rmi\phi}\beta_3 + \rme^{-\rmi\phi}\beta_4 &=
 \rme^{\rmi\phi} \beta_3 - \rme^{\rmi\phi} \beta_4,\qquad
\text{using \eqref{ne:14},} \nonumber \\
\Rightarrow\quad \beta_3 &= -\rmi\cot\phi\; \beta_4,
\end{align}
finally getting
\begin{equation}
  \label{ne:15}
  \begin{aligned}
    \beta_1 &= -\rmi\rme^{-\rmi\phi}\cot\phi\; \beta_4 \\
    \beta_2 &= -\rmi\rme^{-\rmi\phi}\beta_4\\
    \beta_3 &= -\rmi\cot\phi\; \beta_4.
  \end{aligned}
\end{equation}
\subsection{Eigenvalue equation}
The undetermined constant $\beta_4$ will be set by normalisation, but
we first turn to the value of $\phi$.  For this we will need 
explicit formul\ae\ for $W_1,\ldots W_4$.

From (\ref{ne:2}), 
\begin{align}
\|\wplus\|^2  &= \left(\curlyF_B \vec{u}_j + \vec{u}_j\right)^\dag \left(
\curlyF_B\vec{u}_j + \vec{u}_j \right) \nonumber \\
&=  \left( \vec{u}_j^\dag + \vec{u}_j^\dag \curlyF_B^\dag \right)
\left( \curlyF_B\vec{u}_j + \vec u \right) \nonumber \\
&= 2\|\vec u \|^2 + \vec{u}_j^\dag ( \curlyF_B^\dag + \curlyF_B) \vec{u}_j,
\end{align}
and
\begin{equation}
  \label{ne:48}
  \|\wminus\|^2  
= 2\|\vec u \|^2 - \vec{u}_j^\dag ( \curlyF_B^\dag + \curlyF_B) \vec{u}_j,
\end{equation}
so we immediately see that
\begin{equation}
  \label{ne:49}
  W_1 = \|\wplus\|^2 + \|\wminus\|^2 = 4\|\vec{u}_j\|^2 = 8,
\end{equation}
and
\begin{equation}
  \label{ne:50}
  W_3 = \|\wplus\|^2 - \|\wminus\|^2 = 2 \vec{u}_j^\dag (\curlyF_B +
\curlyF_B^\dag ) \vec{u}_j.
\end{equation}

At this point we use the formula \eqref{eq:84} for the entries of $\curlyF_B$ 
to deduce
that the $nm$th entry of the matrix $\curlyF_B+\curlyF_B^\dag$ is
\begin{equation}
  \label{ne:51}
  \frac2{\sqrt{B}} \cos \left( \frac{2\pi(n-1)(m-1)}{B} \right),
\end{equation}
and the $n$th entry of the vector $(\curlyF_B+\curlyF_B^\dag)\vec{u}_j$ is
\begin{multline}
  \label{ne:52}
  \frac2{\sqrt{B}} \cos \left( \frac{2\pi(n-1)(j-1)}{B} \right)
+\frac2{\sqrt{B}} \cos \left( \frac{2\pi(n-1)(B+1-j)}{B} \right)
\\=\frac4{\sqrt{B}} \cos \left( \frac{2\pi(n-1)(j-1)}{B} \right)
\end{multline}
so that
\begin{align}
  \vec{u}_j^\dag ( \curlyF_B +\curlyF_B^\dag ) \vec{u}_j &=
\frac4{\sqrt{B}} \cos \left( \frac{2\pi(j-1)^2}{B} \right)
+\frac4{\sqrt{B}} \cos \left( \frac{2\pi(j-1)(B+1-j)}{B} \right)\nonumber \\
&= \frac8{\sqrt{B}}\cos \left( \frac{2\pi(j-1)^2}B \right).
  \label{ne:53}
\end{align}
This means that, from \eqref{ne:50},
\begin{equation}
  \label{ne:54}
  W_3 = \frac{16}{\sqrt{B}} \cos \left( \frac{2\pi(j-1)^2}{B} \right).
\end{equation}
In a completely analogous way we find that
\begin{equation}
  \label{ne:55}
  \begin{aligned}
    \|\wplusi\|^2  &= 2\|\vec{v}_j\|^2 + \rmi \vec{v}_j^\dag ( \curlyF_B^\dag -
\curlyF_B )\vec{v}_j, \\
    \|\wminusi\|^2  &= 2\|\vec{v}_j\|^2 - \rmi \vec{v}_j^\dag ( \curlyF_B^\dag -
\curlyF_B)\vec{v}_j. 
  \end{aligned}
\end{equation}
Similar to how we arrived at \eqref{ne:53}, 
\begin{equation}
  \label{ne:56}
  \vec{v}_j^\dag ( \curlyF_B^\dag -\curlyF_B ) \vec{v}_j =
- \frac{8\rmi}{\sqrt{B}}\sin \left( \frac{2\pi(j-1)^2}B \right).
\end{equation}
Therefore
\begin{equation}
  \label{ne:57}
  W_2 =  \|\wplusi\|^2 + \|\wminusi\|^2 = 4\|\vec{v}_j\|^2 =  8
\end{equation}
and
\begin{equation}
  \label{ne:58}
    W_4 = \|\wplusi\|^2 - \|\wminusi\|^2 = \frac{16}{\sqrt{B}} \sin\left(
\frac{2\pi(j-1)^2}{B}\right).
\end{equation}

The big matrix in \eqref{ne:12} becomes now
\begin{equation}
  \label{ne:59}
    \left( {\renewcommand{\arraystretch}{1.2} \begin{array}{cccc}
 0 & 0 & 1 & 0 \\
 0 & 0 & 0 & -\rmi \\
 1+\frac{K(\kappa)}2 & -\frac{\rmi K(\kappa)}2 & 
 \frac{K(\kappa)}{\sqrt{B}}\cos(\frac{2\pi(j-1)^2}B)  & 
 \frac{\rmi K(\kappa)}{\sqrt{B}} \sin(\frac{2\pi(j-1)^2}B) \\
 \frac{K(\kappa)}2 & -\rmi-\frac{\rmi K(\kappa)}2 & 
 \frac{K(\kappa)}{\sqrt{B}}\cos(\frac{2\pi(j-1)^2}B)  & 
 \frac{\rmi K(\kappa)}{\sqrt{B}} \sin(\frac{2\pi(j-1)^2}B) 
         \end{array}} \right)
\end{equation}
To calculate its characteristic polynomial we use profitably the Schur
complement formula,
\begin{equation}
  \label{ne:60}
  \det \left( \begin{array}{cc}
                A & B \\ C & D 
              \end{array} \right) = \det A \det(D-CA^{-1}B),
\end{equation}
to deduce that
\begin{align}
  \label{ne:61}
\det&     \left( {\renewcommand{\arraystretch}{1.2} \begin{array}{cccc}
 -\lambda & 0 & 1 & 0 \\
 0 & -\lambda & 0 & -\rmi \\
 1+\frac{K(\kappa)}2 & -\frac{\rmi K(\kappa)}2 & 
 \frac{K(\kappa)}{\sqrt{B}}\cos(\frac{2\pi(j-1)^2}B)-\lambda  & 
 \frac{\rmi K(\kappa)}{\sqrt{B}} \sin(\frac{2\pi(j-1)^2}B) \\
 \frac{K(\kappa)}2 & -\rmi-\frac{\rmi K(\kappa)}2 & 
 \frac{K(\kappa)}{\sqrt{B}}\cos(\frac{2\pi(j-1)^2}B)  & 
 \frac{\rmi K(\kappa)}{\sqrt{B}} \sin(\frac{2\pi(j-1)^2}B) - \lambda 
         \end{array}} \right) \\
&=\lambda^2 \det \left(  \left( 
{\renewcommand{\arraystretch}{1.2} \begin{array}{cc}
 \frac{K(\kappa)}{\sqrt{B}}\cos(\frac{2\pi(j-1)^2}B)-\lambda  & 
 \frac{\rmi K(\kappa)}{\sqrt{B}} \sin(\frac{2\pi(j-1)^2}B) \\
 \frac{K(\kappa)}{\sqrt{B}}\cos(\frac{2\pi(j-1)^2}B)  & 
 \frac{\rmi K(\kappa)}{\sqrt{B}} \sin(\frac{2\pi(j-1)^2}B) - \lambda 
         \end{array}} \right)\right.\nonumber \\ &\quad + \frac1{\lambda} 
\left( \begin{array}{cc}
         1+\frac{K(\kappa)}2 & -\frac{\rmi K(\kappa)}2 \\
         \frac{K(\kappa)}2 & -\rmi-\frac{\rmi K(\kappa)}2 
       \end{array} \right)\left( \begin{array}{cc}
                                     1 & 0 \\
                                   0 & -\rmi
                                 \end{array} \Bigg)
 \right) \nonumber \\
&=\det  \left( 
{\renewcommand{\arraystretch}{1.2} \begin{array}{cc}
 \frac{K(\kappa)\lambda}{\sqrt{B}}\cos(\frac{2\pi(j-1)^2}B)+\frac{K(\kappa)}2
+1-\lambda^2  & 
 \frac{\rmi K(\kappa)\lambda}{\sqrt{B}} \sin(\frac{2\pi(j-1)^2}B)-
\frac{K(\kappa)}2 \\
 \frac{K(\kappa)\lambda}{\sqrt{B}}\cos(\frac{2\pi(j-1)^2}B) + \frac{K(\kappa)}2
  & 
 \frac{\rmi K(\kappa)\lambda}{\sqrt{B}} \sin(\frac{2\pi(j-1)^2}B)
-\frac{K(\kappa)}2 -1 - \lambda^2
         \end{array}} \right)\nonumber
\end{align}
and, continuing,
\begin{align}
  &= \left(\frac{K(\kappa)\lambda}{\sqrt{B}}\cos\left(\frac{2\pi(j-1)^2}B
\right)+\frac{K(\kappa)}2
+1-\lambda^2  \right)\nonumber \\
&\qquad\times\left( \frac{\rmi K(\kappa)\lambda}{\sqrt{B}} \sin\left(\frac{2\pi(j-1)^2}B
\right) -\frac{K(\kappa)}2 -1 - \lambda^2\right) \nonumber \\
&\quad-\left( \frac{\rmi K(\kappa)\lambda}{\sqrt{B}} \sin\left(
\frac{2\pi(j-1)^2}B\right)
-\frac{K(\kappa)}2 \right)\left(
 \frac{K(\kappa)\lambda}{\sqrt{B}}\cos\left(\frac{2\pi(j-1)^2}B\right) 
+ \frac{K(\kappa)}2 \right) \nonumber \\
  &=-(1+\lambda^2) \frac{K(\kappa)\lambda}{\sqrt{B}} \cos\left( 
\frac{2\pi(j-1)^2}{B}\right) + (1-\lambda^2)\frac{\rmi K(\kappa)\lambda}
{\sqrt{B}}\sin \left( \frac{2\pi(j-1)^2}{B} \right) 
\nonumber \\
&\qquad - \frac{K(\kappa)}2(1+ \lambda^2) 
- (1-\lambda^2)(1+\lambda^2) - \frac{K(\kappa)}2(1-\lambda^2) 
\nonumber\\
&=\lambda^4 - \frac{\lambda^3 K(\kappa)}{\sqrt{B}}\rme^{2\pi\rmi(j-1)^2/B}
- \frac{\lambda K(\kappa)}{\sqrt{B}}\rme^{-2\pi\rmi (j-1)^2/B} - 1 -
K(\kappa). 
  \label{ne:62}
\end{align}
The eigenphases $\rme^{\rmi\phi}$ appearing in the solutions \eqref{ne:15} are
the zeros $\lambda$ of the polynomial \eqref{ne:62}.  Note that if
$\kappa=0$, and there is no perturbation, then $K(0)=0$, and the
polynomial \eqref{ne:62} reduces to $\lambda^4-1$ with roots 
$\lambda=\pm1, \pm\rmi$, the unperturbed eigenvalues of $\curlyF_B$, as
to be expected.

\subsection{Explicit form of the perturbed eigenvectors}
The eigenvectors we have found are of the form \eqref{ne:5},
\begin{equation}
  \label{ne:63}
  \vec{f} = \beta_1 \curlyF_B\vec{u}_j + \beta_2 \rmi\curlyF_B\vec{v}_j +
\beta_3 \vec{u}_j + \beta_4 \vec{v}_j,
\end{equation}
where $\beta_1,\ldots,\beta_3$ satisfy \eqref{ne:15} depending on 
$\phi$, the phase of one of the four roots of \eqref{ne:62}.  Putting
these into \eqref{ne:63}, and re-naming $-\rmi\beta_4=Z$, we get
\begin{align}
  \vec{f} &= \beta_4\left( -\rmi\cot\phi (\rme^{-\rmi\phi} \curlyF_B \vec{u}_j +
\vec{u}_j ) + \rme^{-\rmi\phi}\curlyF_B \vec{v}_j + \vec{v}_j)\right) 
\nonumber\\
&= Z\left( \cot\phi (\rme^{-\rmi\phi}\curlyF_B \vec{u}_j+\vec{u}_j) +
\rmi(\rme^{-\rmi\phi}\curlyF_B\vec{v}_j + \vec{v}_j)\right).
  \label{ne:64}
\end{align}
Let
\begin{equation}
  \label{ne:65}
  \begin{aligned}
    \vec{z}_1 &\coloneq \rme^{-\rmi\phi} \curlyF_B\vec{u}_j + \vec{u}_j \\
    \vec{z}_2 &\coloneq \rme^{-\rmi\phi} \curlyF_B\vec{v}_j + \vec{v}_j,
  \end{aligned}
\end{equation}
so that
\begin{equation}
  \label{ne:66}
  \vec{f} = Z \left( \cot\phi\vec{z}_1 + \rmi\vec{z}_2 \right).
\end{equation}
Then
\begin{align}
  \vec{z}_1^\dag \vec{z}_2 &= (\rme^{\rmi\phi}\vec{u}_j^\dag \curlyF_B^\dag + 
\vec{u}_j^\dag) (\rme^{-\rmi\phi}\curlyF_B\vec{v}_j + \vec{v}_j ) \nonumber \\
&= \vec{u}_j^\dag \vec{v}_j + \rme^{\rmi\phi} \vec{u}_j^\dag \curlyF_B^\dag 
\vec{v}_j 
+ \rme^{-\rmi\phi}\vec{u}_j^\dag \curlyF_B \vec{v}_j + \vec{u}_j^\dag\vec{v}_j 
\nonumber \\
&= 0,
  \label{ne:67}
\end{align}
because $\vec{u}_j^\dag\vec{v}_j=0$ by construction, and since $\vec{u}_j^\dag(
\curlyF_B^\dag)^2= \vec{u}_j^\dag$ and $\curlyF_B^2\vec{v}_j=-\vec{v}_j$ we have 
that
\begin{equation}
  \label{ne:68}
  \vec{u}_j^\dag \curlyF_B\vec{v}_j = -\vec{u}_j^\dag\curlyF_B^\dag \vec{v}_j =
- \vec{u}_j^\dag (\curlyF_B^\dag)^2 \curlyF_B \vec{v}_j = - \vec{u}_j^\dag
\curlyF_B\vec{v}_j,
\end{equation}
and similar for $\vec{u}_j^\dag \curlyF_B^\dag \vec{v}_j$.

A consequence of the orthogonality \eqref{ne:67} is that
\begin{equation}
  \label{ne:69}
  \|\vec{f}\|^2 = |Z|^2 \left( \cot^2\phi \|\vec{z}_1\|^2 + \| \vec{z}_2\|^2
 \right).
\end{equation}

From \eqref{ne:65},
\begin{equation}
  \label{ne:73}
  \|\vec{z}_1\|^2 = 2\|\vec{u}_j\|^2 + \vec{u}_j^\dag
(\rme^{-\rmi\phi}\curlyF_B + \rme^{\rmi\phi}\curlyF_B^\dag)\vec{u}_j.
\end{equation}

Similarly to how \eqref{ne:53} was obtained, the $mn$th entry of the
matrix $\rme^{-\rmi\phi}\curlyF_B + \rme^{\rmi\phi}\curlyF_B^\dag$ is
\begin{equation}
  \label{ne:70}
  \frac2{\sqrt{B}} \cos\left( \phi - \frac{2\pi(n-1)(m-1)}{B}\right)
\end{equation}
and the $m$th entry of the vector $(\rme^{-\rmi\phi}\curlyF_B + 
\rme^{\rmi\phi}\curlyF_B^\dag)\vec{u}_j$ is
\begin{multline}
  \label{ne:71}
  \frac2{\sqrt{B}} \cos\left( \phi - \frac{2\pi(j-1)(m-1)}{B}\right)
+\frac2{\sqrt{B}} \cos\left( \phi - \frac{2\pi(B+1-j)(m-1)}{B}\right)\\
=\frac4{\sqrt{B}}\cos\phi \cos\left(\frac{2\pi(j-1)(m-1)}{B}\right),
\end{multline}
leading to
\begin{equation}
  \label{ne:72}
\vec{u}_j^\dag (\rme^{-\rmi\phi}\curlyF_B + 
\rme^{\rmi\phi}\curlyF_B^\dag)\vec{u}_j
=\frac8{\sqrt{B}}\cos\phi \cos\left(\frac{2\pi(j-1)^2}{B}\right),
\end{equation}
and so we get from \eqref{ne:73},
\begin{equation}
  \label{ne:74}
  \|\vec{z}_1\|^2 = 4 +\frac8{\sqrt{B}}\cos\phi \cos\left(\frac{2\pi(j-1)^2}{B}
\right).
\end{equation}

Proceeding similarly,
\begin{equation}
  \label{ne:75}
 \|  \vec{z}_2\|^2 = 2\| \vec{v}_j \|^2 + \vec{v}_j^\dag (
\rme^{\rmi\phi}\curlyF_B^\dag + \rme^{-\rmi\phi}\curlyF_B ) \vec{v}_j,
\end{equation}
and we find that
\begin{equation}
  \label{ne:76}
  \vec{v}_j^\dag( \rme^{\rmi\phi} \curlyF_B^\dag + \rme^{\rmi\phi}\curlyF_B)
  \vec{v}_j = \frac{8}{\sqrt{B}}\sin\phi \sin\left( \frac{2\pi(j-1)^2}{B}
\right),
\end{equation}
so that
\begin{equation}
  \label{ne:77}
  \|  \vec{z}_2\|^2 = 4 + \frac{8}{\sqrt{B}}\sin\phi 
\sin\left( \frac{2\pi(j-1)^2}{B} \right).
\end{equation}

Putting together \eqref{ne:77} and \eqref{ne:74} into \eqref{ne:69} we get
\begin{align}
  \label{ne:78}
  \|\vec{f}\|^2 &= |Z|^2 \bigg( 4\cot^2\phi +\frac8{\sqrt{B}}\cot^2\phi
                 \cos\phi \cos\left(\frac{2\pi(j-1)^2}B \right) 
\nonumber \\ &\qquad 
+ 4 + \frac8{\sqrt{B}} \sin\phi \sin\left(\frac{2\pi(j-1)^2}B \right)  \bigg)
\\
&= |Z|^2 \cosec^2\phi \left( 4 + \frac8{\sqrt{B}}\left( \cos^3\phi
 \cos\left(\frac{2\pi(j-1)^2}B \right) + \sin^3\phi
  \sin \left( \frac{2\pi(j-1)^2}B \right) \right) \right).
\nonumber 
\end{align}

Let us denote $\vec{f}=(f_1,\ldots,f_B)$. We now turn to explicit
calculation of the components squared, $|f_n|^2$.  It is prudent
to consider the cases $n=j, B+2-j$ separately.

Let us note for subsequent calculations, that the $n$th entry of the
vector $\curlyF_B\vec{u}_j$ is
\begin{align}
  \label{ne:79}
  \frac1{\sqrt{B}} \rme^{2\pi\rmi(n-1)(j-1)/B} + \frac1{\sqrt{B}}
\rme^{2\pi\rmi(n-1)(B+1-j)/B} = \frac2{\sqrt{B}} 
  \cos\left( \frac{2\pi(n-1)(j-1)}B \right),
\end{align}
and the $n$th entry of the vector $\curlyF_B\vec{v}_j$ is
\begin{align}
  \label{ne:79a}
  \frac1{\sqrt{B}} \rme^{2\pi\rmi(n-1)(j-1)/B} - \frac1{\sqrt{B}}
\rme^{2\pi\rmi(n-1)(B+1-j)/B} = \frac{2\rmi}{\sqrt{B}} 
  \sin\left( \frac{2\pi(n-1)(j-1)}B \right).
\end{align}

Assume that $n\neq j, B+2-j$. Then from \eqref{ne:64} and the fact that
the $n$th components of $\vec{u}_j$ and $\vec{v}_j$ are zero,
\begin{align}
  f_n &= Z \left( \frac{2\cot\phi}{\sqrt{B}}\rme^{-\rmi\phi}    \nonumber
\cos\left( \frac{2\pi(n-1)(j-1)}B \right) - \frac2{\sqrt{B}}
\rme^{-\rmi\phi} \sin \left(\frac{2\pi(n-1)(j-1)}B \right) \right) \\
&= \frac{2Z}{\sqrt{B}}\rme^{-\rmi\phi} \cosec\phi \cos \left( 
\phi+\frac{2\pi(n-1)(j-1)}B \right),
  \label{ne:80}
\end{align}
and
\begin{equation}
  \label{ne:81}
  |f_n|^2 = \frac{4|Z|^2}B  \cosec^2\phi \cos^2 \left( \phi+
\frac{2\pi(n-1)(j-1)}B \right).
\end{equation}

If $n=j$ or $n=B+2-j$, then $f_n$ contains contributions from the vectors
$\vec{u}_j$ and $\vec{v}_j$, giving
\begin{align}
  f_n &= Z\Bigg( \frac{2\cot\phi}{\sqrt{B}} \rme^{-\rmi\phi}  
\cos \left( \frac{2\pi(n-1)(j-1)}{B} \right) + \cot\phi \nonumber
\\ &\qquad - \frac2{\sqrt{B}} \rme^{-\rmi\phi} \sin \left( \frac{2\pi(n-1)
(j-1)}B \right) \pm \rmi \Bigg) \nonumber \\
&= Z\cosec\phi \left( \frac2{\sqrt{B}} \rme^{-\rmi\phi}
  \cos\left( \phi + \frac{2\pi(n-1)(j-1)}{B}\right) + \rme^{\pm\rmi\phi} 
\right)
  \label{ne:82}
\end{align}
with the ``$+$'' sign taken for $n=j$ and the ``$-$'' sign for $n=B+2-j$.
It follows that
\begin{multline}
  \label{ne:96}
|f_n|^2 = |Z|^2\cosec^2\phi \left( \frac2{\sqrt{B}} \rme^{-\rmi\phi}
  \cos\left( \phi + \frac{2\pi(n-1)(j-1)}{B}\right) + \rme^{\pm\rmi\phi} 
 \right) \\
\times \left( \frac2{\sqrt{B}} \rme^{\rmi\phi}
  \cos\left( \phi + \frac{2\pi(n-1)(j-1)}{B}\right) + \rme^{\mp\rmi\phi} 
 \right).
\end{multline}
This becomes, for $n=j$,
\begin{equation}
\label{ne:83}
  |f_j|^2 = |Z|^2\cosec^2\phi \left( \frac4B \cos^2 \left( \phi
      + \frac{2\pi(j-1)^2}{B} \right) 
    + \frac{4}{\sqrt{B}}\cos2\phi \cos \left( \phi + \frac{2\pi(j-1)^2}{B}
 \right)  + 1\right)
\end{equation}
and for $n=B+2-j$,
\begin{equation}
  \label{ne:83a}
  |f_{B+2-j}|^2 = |Z|^2\cosec^2\phi \left( \frac4B \cos^2 \left( \phi
      - \frac{2\pi(j-1)^2}{B} \right) 
    + \frac{4}{\sqrt{B}} \cos \left( \phi - \frac{2\pi(j-1)^2}{B}
 \right)  + 1\right)
\end{equation}

It is possible to verify the calculation of the norm $\|\vec{f}\|^2$,
\eqref{ne:78}
by summing over \eqref{ne:81} with \eqref{ne:83} and \eqref{ne:83a}, but 
we do not do that here.  Moreover, the components \eqref{ne:81},
\eqref{ne:83} and \eqref{ne:83a}, together with \eqref{ne:78}, are sufficient
information to calculate the entropy $\ent(\vec{f})$ \emph{exactly}.  

\subsection{Entropy calculation}
We now focus on the calculation of the entropy up to terms small in
the size $B$ of the matrix.

From \eqref{ne:78}, we have
\begin{equation}
  \label{ne:97}
  \| \vec{f} \|^2 = 4|Z|^2 \cosec^2\phi \left( 1 + \Ord(B^{-1/2}) \right).
\end{equation}
We also deduce from \eqref{ne:83} and \eqref{ne:83a} that
\begin{equation}
  \label{ne:98}
  \begin{aligned}
    |f_j|^2 &= |Z|^2\cosec^2\phi \left( 1 + \Ord(B^{-1/2} \right), \\
    |f_{B+2-j}|^2 &= |Z|^2\cosec^2\phi \left( 1 + \Ord(B^{-1/2} \right).
  \end{aligned}
\end{equation}
This means that
\begin{equation}
  \label{ne:99}
  \frac{|f_j|^2}{\|\vec{f}\|^2} = \frac14 + \Ord(B^{-1/2}),
\end{equation}
and similarly for $f_{B+2-j}$.

If $h(x) \coloneq -x\log x$, so that
\begin{equation}
  \label{ne:100}
  \ent(\vec{f}) = \sum_{n=1}^B h\left( \frac{|f_n|}{\|\vec{f}\|^2} \right),
\end{equation}
since $h$ is differentiable on $(0,1)$, 
\begin{equation}
  \label{ne:101}
  h\left( \frac{|f_j|}{\|\vec{f}\|^2}\right)
 = h({\textstyle\frac14}) + \Ord(B^{-1/2}) 
= \frac12\log 2 + \Ord(B^{-1/2}),
\end{equation}
and $f_{B+2-j}$ contributes the same.  

The main contribution to $\ent(f)$ comes from the remaining components,
with $n\neq j, B+2-j$.  Coming from \eqref{ne:81} we get
\begin{equation}
  \label{ne:105}
  \frac{|f_n|^2}{\|\vec{f}\|^2} = \frac1{B}
\cos^2\left(\phi + \frac{2\pi(n-1)(j-1)}{B} \right) 
+\Ord(B^{-3/2}).
\end{equation}
The contribution of each of these components to the entropy is
\begin{multline}
  \label{ne:106}
  h\left( \frac{|f_n|^2}{\|\vec{f}\|^2} \right) = 
\frac{\log B}{B}\cos^2\left(\phi + \frac{2\pi(n-1)(j-1)}{B} \right) \\
- \frac1B \cos^2\left(\phi + \frac{2\pi(n-1)(j-1)}{B} \right) 
\log \left(\cos^2\left(\phi + \frac{2\pi(n-1)(j-1)}{B} \right)\right) \\
+ \Ord\left( \frac{\log B}{B^{3/2}}\right).
\end{multline}

Adding up the contributions we have
\begin{equation}
  \label{ne:107}
  \ent(\vec{f}) = S_1(\phi,B)\log B - S_2(\phi,B) + 2 \frac12\log2+
\Ord\left( \frac{\log B}{B^{1/2}} \right),
\end{equation}
where
\begin{equation}
  \label{ne:108}
  S_1(\phi, B)\coloneq \frac1B \sum_{\substack{n=1\\n\neq j,B+2-j}}^B
 \cos^2 \left( \phi+ \frac{2\pi(n-1)(j-1)}{B} \right)
\end{equation}
and
\begin{equation}
  \label{ne:109}
    S_2(\phi,B)\coloneq \frac1B \sum_{\substack{n=1\\n\neq j,B+2-j}}^B
 \cos^2 \left( \phi+ \frac{2\pi(n-1)(j-1)}{B} \right)
\log \left(  \cos^2 \left( \phi+ \frac{2\pi(n-1)(j-1)}{B} \right)
\right).
\end{equation}
The quantities $S_1(\phi,B)$ and $S_2(\phi,B)$ are Riemann sums and can
be estimated up to errors of order $\Ord(B^{-1})$ by integrals:
for $S_1$,
\begin{align}
  S_1(\phi,B) &= \int_0^1 \cos^2(\phi+2\pi(j-1)x)\,\rmd x + \Ord(B^{-1}) 
\nonumber \\
&= \frac1{2\pi} \int_0^{2\pi} \cos^2(\phi+y)\,\rmd y + \Ord(B^{-1})
\nonumber \\
&=\frac12 + \Ord(B^{-1}),  
  \label{ne:110}
\end{align}
and
\begin{align}
  S_2(\phi,B) &= \int_0^1 \cos^2(\phi+2\pi(j-1)x) \log(\cos^2(\phi+2\pi(j-1)x))
\,\rmd x +\Ord(B^{-1})
\nonumber \\
&=\frac{1}{2\pi} \int_0^{2\pi} \cos^2y \log(\cos^2 y)\,\rmd y +\Ord(B^{-1})
\nonumber \\
&=\frac1{2\pi} \int_0^{2\pi} \left({\textstyle \frac12+ \frac12\cos2y}\right)
\log \left({\textstyle \frac12+ \frac12\cos2y}\right)\,\rmd y +\Ord(B^{-1}).
  \label{ne:111}
\end{align}
The integral in \eqref{ne:111} is of the type considered in appendix
\ref{app:D}.  From the evaluation contained therein, we find that
\begin{equation}
  \label{ne:112}
  S_2(\phi,B) = -\log2 + \frac12 + \Ord(B^{-1}).
\end{equation}
Finally, putting \eqref{ne:112} and \eqref{ne:110} into \eqref{ne:107},
the entropy of $\vec{f}$ is
\begin{equation}
  \label{ne:113}
  \ent(\vec{f}) = \frac12\log B + 2\log 2 - \frac12
+ \Ord\left( \frac{\log B}{B^{1/2}} \right).
\end{equation}
\section{Evaluation of an  integral} \label{app:D}
In this appendix we evaluate the integral
\begin{equation}
  \label{ne:84}
  I(\alpha,\beta,a,b) \coloneq \int_0^{2\pi} (\alpha + \beta\cos x)
  \log\left( a + b\cos x \right)\, \rmd x,\qquad a>0, |b|<a.
\end{equation}
In the application we have in mind $\alpha=a$ and $\beta=b$, but it
is no extra effort to consider the more general form.

For simplicity we initially suppose that $a=1+t^2$, $b=2t$ for some
$t$ with $|t|<1$.  Then
\begin{align}
  \label{ne:85}
  \log(1+t^2 + 2t\cos x) &= \log \left( (1+t\rme^{\rmi x})
(1+t\rme^{-\rmi x}) \right) \\
&= \log(1+t\rme^{\rmi x}) + \log(1+t \rme^{-\rmi x}).
\label{ne:89}
\end{align}
So
\begin{equation}
  \label{ne:86}
  I(\alpha, \beta, 1+t^2, 2t) = \int_0^{2\pi} \left(\alpha + \frac{\beta}2
(\rme^{\rmi x} + \rme^{-\rmi x})\right) 
\left(  \log(1+t\rme^{\rmi x}) + \log(1+t\rme^{-\rmi x} )\right)\,\rmd x.
\end{equation}
We write the logarithmic terms as absolutely uniformly (in $x$) convergent
series:
\begin{equation}
  \label{ne:87}
  \log(1+t\rme^{\pm\rmi x}) = -
\sum_{n=1}^\infty \frac{(-t)^n \rme^{\pm\rmi n x}}n,
\end{equation}
Inserting \eqref{ne:87} into \eqref{ne:86} we find that only the
$n=1$ terms of the summations contribute and we get
\begin{equation}
  \label{ne:88}
  I(\alpha, \beta, 1+t^2, 2t) = 2\pi\beta t.
\end{equation}

If $a$ and $b$ happen not to be of the special form above, we can write
\begin{multline}
  \label{ne:90}
  \log(a+b\cos x) = \log \left( \frac{2a}{b^2} (a-\sqrt{a^2-b^2})
    + \frac2b(a-\sqrt{a^2-b^2})\cos x \right) \\- \log\left(
  \frac2{b^2}(a-\sqrt{a^2-b^2})\right),
\end{multline}
where the $x$-dependent term \emph{is} of the form \eqref{ne:85}, with 
\begin{equation}
  \label{ne:91}
  t = \frac1b(a-\sqrt{a^2-b^2})
\end{equation}
and
\begin{align}
  1+ t^2 &= 1 + \frac1{b^2}(a-\sqrt{a^2-b^2})^2 \nonumber \\
  &= \frac{2a}{b^2}(a-\sqrt{a^2-b^2}).
  \label{ne:92}
\end{align}
Thus,
\begin{equation}
  \label{ne:93}
  I(\alpha,\beta,a,b) = I(\alpha,\beta,1+t^2,2t) -
2\pi\alpha \log\left(
  \frac2{b^2}(a-\sqrt{a^2-b^2})\right)
\end{equation}
with $t$ given by \eqref{ne:91}.  Noting that
\begin{align}
  - \log\left(
  \frac2{b^2}(a-\sqrt{a^2-b^2})\right)  &= \log \left( \frac{b^2}2
\frac1{a-\sqrt{a^2-b^2}} \right) \nonumber \\
&= \log \left( \frac{a+\sqrt{a^2-b^2}}2 \right),
  \label{ne:94}
\end{align}
we get that
\begin{equation}
  \label{ne:95}
  I(\alpha,\beta,a,b) = 2\pi\alpha \log \left( \frac{a+\sqrt{a^2-b^2}}{2}
 \right) + 2\pi\beta \left( \frac{a-\sqrt{a^2-b^2}}{b} \right).
\end{equation}
The case $\alpha=1, \beta=0$ of this integral is formula
\textbf{4.224.9} of \cite{gra:tis}.

Finally, if $\alpha=a$ and $\beta=b$, then the integrand \eqref{ne:84}
is bounded and we may increase $b$ to $a$ giving
\begin{equation}
\label{eq:33}
I(a,a,a,a) = 2\pi a \left(1+\log\left(\frac{a}2\right) \right).
\end{equation}

\end{document}